\documentclass{article}

\usepackage{jcappub} % for details on the use of the package, please
\usepackage[utf8]{inputenc}
\usepackage{appendix}
\usepackage{lineno}
\usepackage{subfigure}
\usepackage{comment}
\usepackage{float}
\usepackage[sort&compress]{natbib}
\usepackage{physics}
\usepackage{xcolor}
\usepackage[normalem]{ulem}

\newcommand{\vect}[1]{\boldsymbol{\mathbf{#1}}}
\newcommand{\unit}[1]{\,\mathrm{#1}}

\usepackage[all]{hypcap} % Sergio: Added this so that figure references jump to image instead of caption

\title{Radio Emission from Atmosphere-Skimming Cosmic Ray Showers in High-Altitude Balloon-Borne Experiments}

\author[a]{Matías Tueros,}
\author[b]{Sergio Cabana-Freire,}
\author[b]{Jaime Álvarez-Muñiz}

\emailAdd{tueros@fisica.unlp.edu.ar}
\emailAdd{sergio.cabana.freire@usc.es}
\emailAdd{jaime.alvarez@usc.es}

\affiliation[a]{Instituto de Fisica La Plata, CONICET-UNLP,
   Diagonal 113 entre 63 y 64 , La Plata, Argentina}
\affiliation[b]{Instituto Galego de Física de Altas Enerxías (IGFAE), Universidade de Santiago de Compostela, 15782
Santiago de Compostela, Spain}

\date{\today}

\begin{document}

\abstract{ 
Atmosphere-skimming air showers are initiated by cosmic rays with incoming directions that allow the cascade to develop entirely within the atmosphere, without reaching the ground. Radio pulses induced by this type of showers have already been observed in balloon-borne experiments such as ANITA, but a detailed characterisation of their properties is lacking. The extreme range of densities in which these cascades can develop gives rise to a wide range of shower profiles, with radio emission characteristics that can differ significantly from those of regular downward-going showers. In this work, we have used the ZHAireS-RASPASS program to characterise the expected radio emission from atmosphere-skimming air showers and its properties. We have studied the interplay between the magnetic field and atmospheric density profile in the expected radio signal, focusing on its detection aboard balloon-borne experiments. The almost horizontal geometry of the events gives rise to a significant \textit{refractive} asymmetry in the spatial distribution of the electric field, due to the propagation of the radio signals across a gradient of index of refraction. In addition, a unique \textit{coherence} asymmetry appears in the intensity of the signals, as a consequence of the cumulative effect of the Earth's magnetic field over the very long distances that these particle cascades traverse. The implications of the peculiar characteristics of the emission are discussed regarding their impact both on the interpretation of collected data and in the exposure of balloon-borne experiments.
}

\maketitle

\section{Introduction}
\label{sec:Intro}

Atmosphere-Skimming (AS) air showers are particle cascades initiated by cosmic rays (and potentially other primary particles such as photons or neutrinos) with trajectories such that the entire development of the cascade takes place inside the atmosphere, without the shower core hitting the ground. This type of events has been identified by the ANtarctic Impulsive Transient Antenna (ANITA) balloon-borne detector \cite{ANITA, ANITAI-II, ANITAIII, ANITAIV}, that collected several nanosecond-long radio pulses with an incoming direction and polarization consistent with the expectation for air showers with AS trajectories\footnote{2, 1, 2 and 2 atmosphere-skimming events were detected in the ANITA I, II, III and IV flights, respectively}. 

Atmosphere-Skimming air showers can be potentially detected with different techniques in current and future experiments. These include observing the electromagnetic emission in radio frequencies using antennas carried in balloon-borne payloads as in the Payload for Ultrahigh Energy Observations (PUEO) \cite{PUEO} (the successor of ANITA), and the Probe Of Extreme Multi-Messenger Astrophysics POEMMA-Balloon with Radio \cite{Olinto:2023vmx}; arrays of radio antennas placed in mountains such as the Beamforming Elevated Array for COsmic Neutrinos (BEACON) \cite{BEACON} or the Giant Radio Array for Neutrino Detection (GRAND) \cite{GRAND} and, for favourable geometries, in flat arrays such as the radio detector extension of the Pierre Auger Observatory \cite{PierreAuger_Radio};  fluorescence-light detectors in ground-based experiments (in a limited range of geometries) such as the Pierre Auger Observatory \cite{PierreAuger_FD, PierreAuger_upward, PierreAuger_HEAT, AugerPrime} and the Telescope Array \cite{TelescopeArray}; or directly observing fluorescence and Cherenkov photons aboard satellite- and balloon-borne experiments such as POEMMA \cite{POEMMA} and the Extreme Universe Space Observatory on a Super Pressure Balloon (EUSO-SPB2) \cite{Eser:2023lck, Adams:2024gsj}, that has already observed events consistent with AS trajectories \cite{Cummings:2023ypo}.

The distinct geometry of atmosphere-skimming showers and the interplay between low atmospheric density and the magnetic field in the evolution of the particle cascade, have a strong impact on the properties of the shower, and on the associated emission in radio frequencies \cite{Glaser:2016qso, Ammerman-Yebra:2023rhr, Tueros_Radio_ICRC2023, Chiche:2024yos}, optical and UV bands \cite{Cummings:2021bhg, Cummings:2023ypo}. 

\begin{figure}[t]
    \centering
    \includegraphics[width = 0.85\textwidth]{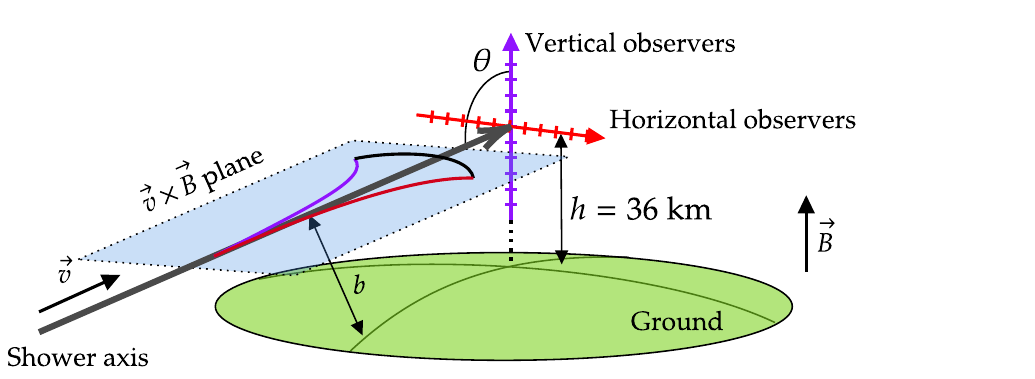} \\
    \includegraphics[width = 0.85\textwidth]{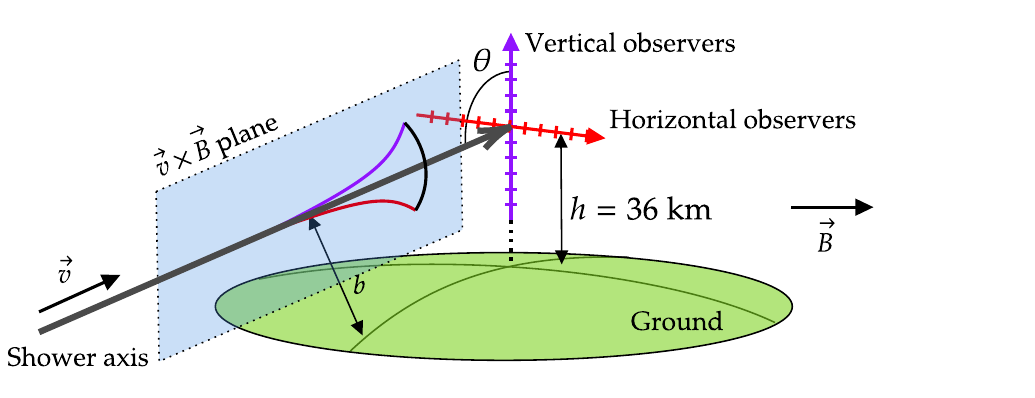}
    \caption{Sketch of the axis of an Atmosphere-Skimming shower (black solid arrow) above the surface of the Earth (in green), its zenith angle $\theta$ and impact parameter $b$ (defined as the distance of closest approach of shower axis to Earth). The observers in the simulations of radio emission performed with ZHAireS-RASPASS are placed along a line parallel (\textit{Horizontal}, in red) and perpendicular (\textit{Vertical}, in blue) to the ground plane with center at the height above sea level of the detector, $h=36\unit{km}$. The antenna positions are contained in the plane perpendicular to the projection of the shower axis on the ground. For illustration purposes, in light blue color we represent the plane where the $\vect{v}\cross\vect{B}$ vector lays, with $\vect{v}$ the direction of the shower axis, for the cases of a vertical magnetic field $\vect{B}$ perpendicular to ground (top panel) and a horizontal magnetic field $\vect{B}$ parallel to ground (bottom panel).}
    \label{fig:RASPASSAntennas}
\end{figure}

Assuming a spherically symmetric earth and atmosphere, the geometry of the shower axis can be uniquely specified with the impact parameter $b$, defined as the minimum distance between the shower axis and the ground, as sketched in figure\,\ref{fig:RASPASSAntennas}. 
In general, atmosphere-skimming particle cascades develop along atmospheric densities much smaller than those relevant for showers whose core hits the ground. For large enough $b$, the limited matter available for development can give rise to particle cascades propagating towards decreasing densities and, in some cases, cascades that could escape the atmosphere without completing their development. 
In addition, the ample distances traveled under the effect of the geomagnetic field can produce a significant spread of the shower front along the direction of the Lorentz force \cite{Cocconi_magnetic}, especially for showers with large $b$ that develop over hundreds of kilometers in a very rarefied atmosphere.

%-------------------------------------------------
%---------------------------------------------
% Sergio: Scheme of antenna placement and directions (preliminar)
%----------------------------------------------------------------------------------
\looseness = -1
The absence of full 4D Monte Carlo simulation tools capable of handling the incoming directions of AS events and calculating their associated electromagnetic emission has limited the progress in this area until very recently, with the release of the ZHAireS-RASPASS \cite{Tueros_RASPASS_ARENA2022} simulation suite.
This allowed us to study the characteristics of the longitudinal (along shower axis) and lateral (perpendicular to shower axis) development of AS showers in \cite{Tueros_Showers_ICRC2023} and \cite{Tueros_Showers_JCAP}.
In this article we will focus on the properties of the emission of AS showers in radio frequencies (MHz$-$ GHz). However, and for self-containment of this article, in section \ref{sec:RASPASS} we overview the main characteristics of AS showers, together with a brief description of the ZHAireS-RASPASS Monte Carlo program. In section \ref{sec:LDF} we discuss the features of radio emission in simulated AS showers, focusing on the geometries and frequency ranges relevant for the case of balloon-borne antenna payloads. The impact of the distinct characteristics of the radio emission on the interpretation of the collected data and their effect on the aperture of high-altitude radio-detectors is addressed in section \ref{sec:implications}. 
The results in this section are relevant for the energy reconstruction of the events recorded with the ANITA payload \cite{ANITAI-II, ANITAIII, ANITAIV} and the determination of its sensitivity to this type of events, as well as in future experiments using the radio technique such as PUEO \cite{PUEO} and the POEMMA-Balloon with Radio \cite{Olinto:2023vmx}. Finally, in section \ref{sec:Conclusions} we summarise and conclude this work. 

%%%%%%%%%%%%%%%%%%%%%%%%%%%%%%%%%%%%%%%%%%%%%%%%%%%%%%%%%%%%%%%%
\section{Features of Atmosphere-Skimming showers relevant for radio emission}
\label{sec:RASPASS}

An alternative description of the geometry of the axis of an AS shower can be done in terms of the incoming zenith angle $\theta$ seen by a detector at an altitude $h$ above sea level, as shown in figure \ref{fig:RASPASSAntennas}. This is equivalent to the description based on the impact parameter $b$ alone, but is better suited for studies regarding detection in high-altitude experiments.
From the point of view of a detector placed at an altitude $h$ above sea level, as the angle $\theta$ increases beyond $90^\circ$ showers become more inclined with respect to the horizontal at the detector (figure \ref{fig:RASPASSAntennas}) and propagate through increasingly denser layers of the atmosphere. Showers find an increasing amount of matter available for development, up to a limiting zenith angle for which the impact parameter $b$ becomes zero and the shower is no longer atmosphere-skimming. 
For the case of a balloon-borne detector hovering at a typical altitude of $h\simeq 36\unit{km}$ a.s.l., the amount of matter available for shower development between the upper limit of the atmosphere and the detector ranges between $\lesssim 200\unit{g/cm^2}$ for showers with $\theta \leq 90^\circ$ and $6.5\times10^4\unit{g/cm^2}$ for the extreme case of a shower grazing the ground at $\theta\simeq 96^\circ$, the angle at which a detector at this altitude sees the horizon, corresponding to showers with $b=0\unit{km}$. Further details can be found in Fig.\,2 of \cite{Tueros_Showers_JCAP}.

\begin{figure}[htb]
    \centering
    \includegraphics[width = .97\textwidth]{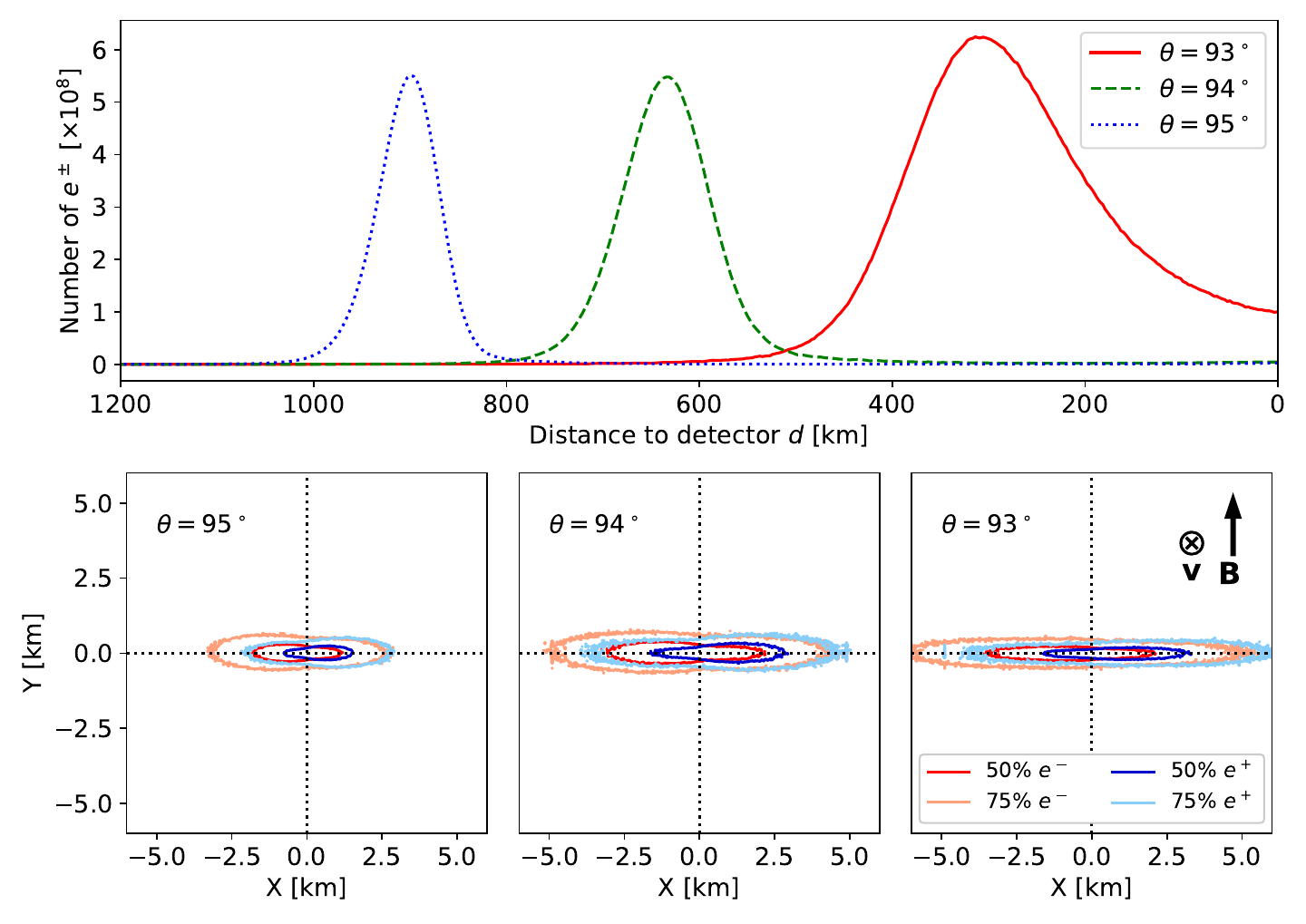}
    \caption{Longitudinal (top panel) and lateral (bottom panels) development of the number of $e^-$ and $e^+$ in proton-induced showers of energy $E_0=10^{18}\unit{eV}$, for zenith angles $\theta=93^\circ,\,94^\circ$ and $95^\circ$ passing at an altitude $h=36\unit{km}$ (figure \ref{fig:RASPASSAntennas}). 
    The longitudinal dimension is measured in terms of the distance $d$ to the detector that is located at $d=0\unit{km}$ by definition. The lateral distribution is shown in a plane perpendicular to the shower axis (with the shower direction $\vect{v}$ \textit{entering the figure} as sketched in the bottom right panel) at the position of the corresponding $X_\mathrm{max}$ in all cases ($\simeq 799\unit{g/cm^2}$, $768\unit{g/cm^2}$ and $791\unit{g/cm^2}$  for $\theta = 93^\circ-94^\circ-95^\circ$ respectively). The showers were simulated using a magnetic field of $50\unit{\mu T}$ perpendicular to the ground (almost parallel to the $Y$ direction as sketched in the bottom right panel).}
    \label{fig:LongLatDev_Bfield}
\end{figure}

The low atmospheric density where AS showers develop, together with the strong dependence of the density profile on the zenith angle, significantly impact the characteristics of the air shower that determine radio emission. One of the most important effects is the dramatic change of the distance from the detector to the position of shower maximum $X_\mathrm{max}$ for a small variation of the shower zenith angle. Since a significant fraction of the total radio emission is expected to come from $X_\mathrm{max}$, this largely determines the attenuation of the signal and hence its intensity at the position of the detector.
As can be seen in the top panel of figure \ref{fig:LongLatDev_Bfield}, the distance between $X_\mathrm{max}$ and a detector at $h=36\unit{km}$ a.s.l. is $\sim 300\unit{km}$ for showers with $\theta = 93^\circ$ and $\sim900\unit{km}$ for $\theta = 95^\circ$.
Moreover, event-to-event fluctuations of a few $\unit{g/cm^2}$ in $X_\mathrm{max}$ can translate into differences of tens of $\unit{km}$ in the distance to detector for showers developing in a very rarefied atmosphere \cite{Tueros_Showers_JCAP}.

The longitudinal and lateral dimensions of the shower are known to determine many properties of its radio emission 
\cite{Zas:1991jv, Alvarez-Muniz:2022uey, Ammerman-Yebra:2023rhr}. 
In the case of AS cascades, their longitudinal spread is significantly larger than in downward-going air showers, extending to approximately $\sim 200\unit{km}$ due to the increased distances the particle cascade needs to travel to find sufficient matter for its development, as illustrated in the top panel of figure \ref{fig:LongLatDev_Bfield} (for further examples the reader is referred to \cite{Tueros_Showers_JCAP}).
Also, as the shower develops, the lateral dimensions of the particle cascade are strongly affected by the interplay between the low density atmosphere and the geomagnetic field, producing a significant spread of the shower front along the direction of the Lorentz force \cite{Tueros_Showers_JCAP}. This is illustrated in the bottom panels of figure \ref{fig:LongLatDev_Bfield}, where the showers are shown to \textit{flatten} in the direction of the geomagnetic force, i.e. perpendicularly to the shower axis $\vect{v}$ and the magnetic field $\vect{B}$. Another effect that is relevant to radio emission is the change in the flow of energy from the hadronic to the electromagnetic component of the shower, since electrons and positrons are responsible for the bulk of the radio emission. The energy flow is driven by $\pi^0$ decay, whose critical energy is influenced by the low atmospheric densities where particle interaction and decay occur, inducing an additional dependence of the radio emission on the shower geometry \cite{Tueros_Showers_JCAP}. The influence of these effects on radio emission is explored in sections \ref{sec:LDF} and \ref{sec:implications} using ZHAireS-RASPASS.  

%-------------------------------------------------
%-------------------------------------------------

ZHAireS-RASPASS, or in short \textbf{R}ASPASS stands for \textbf{A}ires \textbf{S}pecial \textbf{P}rimary for \textbf{A}tmospheric \textbf{S}kimming \textbf{S}howers. It was created in 2011 as an extension module within the Aires shower Monte Carlo simulation program \cite{AIRES}. Its main purpose was to simulate AS showers in response to the detection of the first ANITA above-horizon events. Over time, RASPASS underwent significant evolution and eventually transformed into a standalone version of ZHAireS (Aires with radio emission calculation capabilities \cite{ZHAireS}). The program now encompasses various modifications to accommodate simulations of diverse shower geometries, including downward-going, upward-going, Earth-skimming, and atmosphere-skimming showers. RASPASS uses physics algorithms identical to the standard Aires and ZHAireS programs. It also adopts the user-friendly input interface from Aires, allowing for seamless integration into existing workflows. 
On the other hand, it also inherits some of the ZHAireS and Aires limitations, of which the most relevant for radio emission are 
the utilization of a single-layer exponential for the modeling of the index of refraction and the assumption of a constant magnetic field intensity and orientation along the whole shower development. Straightforward calculations using the International Geomagnetic Reference Field IGRF13 model \cite{IGRF13} indicate that, for the range of shower geometries studied in this work and for an experiment situated at the South Pole, the variations in magnetic force intensity are less than 15\% over the hundreds of kilometers spanning the cascade's development, while the change in the geomagnetic angle is well below $10^\circ$. Consequently, we anticipate that assuming a constant magnetic field should not significantly impact the main properties of the radio emission in AS showers described here. In other locations, where the geomagnetic field gradient is smaller, this approximation is expected to be even more accurate.

%%%%%%%%%%%%%%%%%%%%%%%%%%%%%%%%%%%%%%%%%%%%%%%%%%%%%%%%%%%%%%%%

\section{Characteristics of the radio signal in Atmosphere-Skimming showers}
\label{sec:LDF}

The electromagnetic emission at radio frequencies ($\mathrm{MHz-GHz}$) in particle cascades is known to be produced mainly through two mechanisms: the \textit{geomagnetic mechanism}, due to the electrical current induced by the deflection of charged particles in the Earth's magnetic field \cite{Kahn-Lerche:1966}; and the \textit{Askar'yan mechanism}, consisting on the development of an excess charge in the particle cascade \cite{Askaryan:1962} due to the incorporation of electrons from the medium to the shower front via Bhabha, Möller and Compton scattering as well as positron annihilation. For showers in air, and especially for AS showers developing in very low density layers of the atmosphere, the contribution of the Askar'yan mechanism is typically expected to be much smaller than the geomagnetic emission \cite{PierreAuger:2014ldh, Schellart:2014oaa}. 
In fact, the radio pulses collected by ANITA with incoming directions consistent with AS showers exhibited the polarization signatures expected for the geomagnetic emission, with the electric field primarily polarized perpendicular to the local magnetic field \cite{ANITAI-II, ANITAIII, ANITAIV}.

\looseness=-1
Motivated by the detection of AS events in the ANITA flights, and also by the development of its successor PUEO \cite{PUEO} as well as by other initiatives such as the POEMMA - Balloon with Radio \cite{Olinto:2023vmx}, in this work we have performed simulations of the radio emission in AS showers focusing on geometries relevant for the case of balloon-borne detectors. In particular, we have considered proton-induced showers with their axes passing at an altitude $h=36\unit{km}$ above sea level and with different zenith angles $\theta$ (and thus different impact parameter $b$, see figure\,\ref{fig:RASPASSAntennas}). Since the bulk of electromagnetic emission in particle cascades is produced by electrons and positrons close to the position of $X_\mathrm{max}$, 
we have only considered the most favourable geometries for radio detection, i.e., for which the shower is fully developed before reaching the detector. For the case of showers passing at $h=36\unit{km}$, this restricts the shower zenith angle to $93^\circ<\theta<96^\circ$ (see figure 2 in \cite{Tueros_Showers_JCAP}). 

%%%%%%%%%%%%%%%%%%%%%%%%%%%%%%%%%%%%%%%%%
\subsection{Radio pulses in the time domain}

\begin{figure}[htb]
    \centering
    \includegraphics[width = \textwidth]{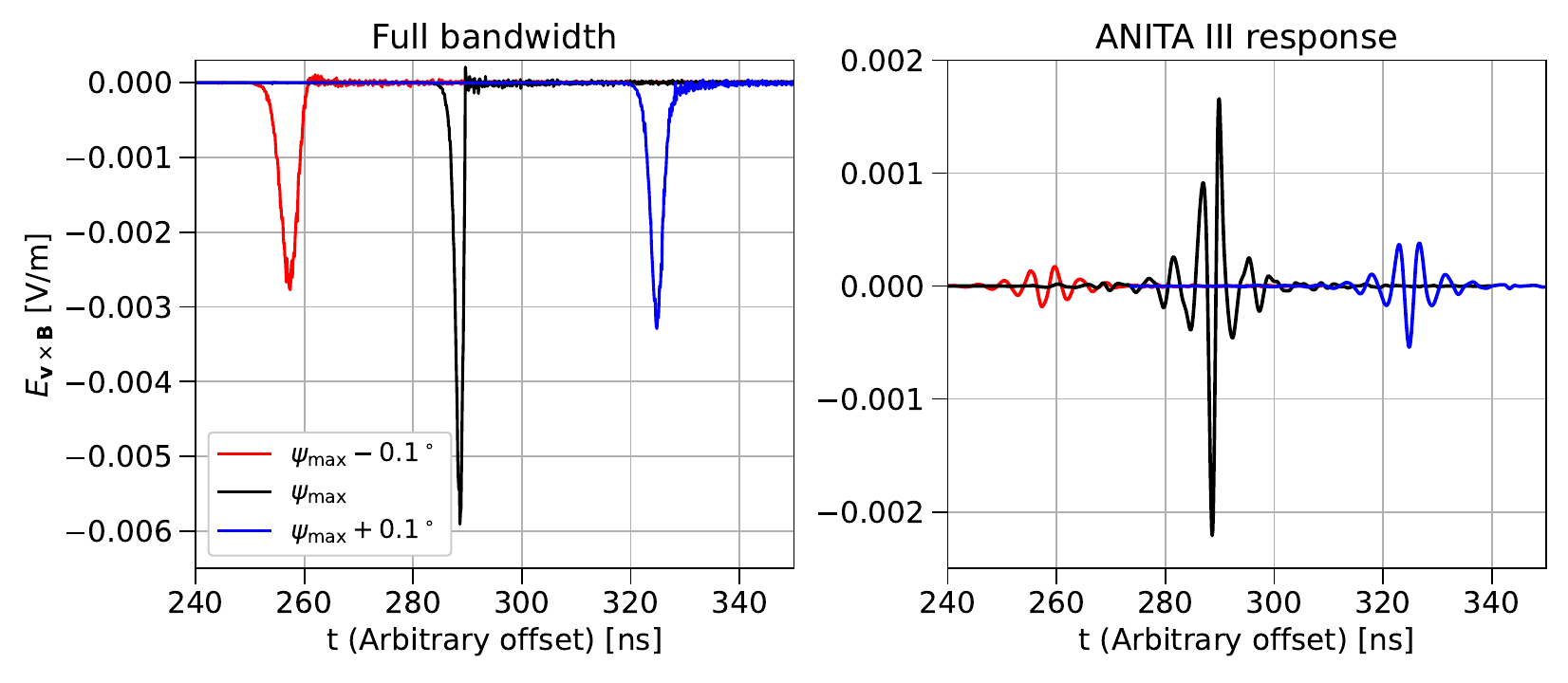}
    \caption{Simulated electric field in the $\vect{v}\times\vect{B}$ direction produced by a proton shower of energy $E_0 = 10^{19.5}\unit{eV}$ passing at an altitude $h=36\unit{km}$ with zenith angle $\theta=94^\circ$. The shower propagates in the presence of a vertical magnetic field of $50\unit{\mu T}$ as sketched in the top panel of figure \ref{fig:RASPASSAntennas}. The electric field is shown at three different positions along the horizontal direction (red axis in the top panel of figure \ref{fig:RASPASSAntennas}), seeing the shower maximum with different angles relative to the off-axis angle $\psi_{\rm max}$ where the coherence is maximal (see text). Left panel: Unfiltered simulation output with a time resolution of $0.1 \unit{ns}$. Right panel, electric fields convoluted with the approximate response of the ANITA III instrument (sensitive to frequencies in the $\sim 200-1200\unit{MHz}$ range).}
    \label{fig:PulsesTimeDomain}
\end{figure}

The coherence properties of the electric field pulses depend strongly on the relative time delays between emissions from all locations of the evolving shower, as seen at the position of the observer. Since most radiation sources (particles) are located near shower maximum, the relative position of $X_{\rm max}$ to the observer determines, to a large extent, the time delays and thus the frequency up to which the Fourier components of the signal add up coherently. When the observer sees shower maximum under a certain angle $\psi_{\rm max}$ with respect to shower axis, such that the relative time delays between emissions from that region are minimal, the received signal is expected to be maximally coherent. For an homogeneous atmosphere, this angle is akin to the well-known Cherenkov angle $\psi_c$, defined as $\cos\psi_c=1/n$ where $n$ is the refractive index of the atmosphere.
In figure \ref{fig:PulsesTimeDomain}, we show an example of a radio pulse obtained with RASPASS at $\psi_{\rm max}$, exhibiting a large peak value and a narrow width typical of a coherent radio emission.
The pulses are also shown at $\psi =\psi_{\rm max}\pm\,0.1^\circ$. In the left panel of figure \ref{fig:PulsesTimeDomain}, the full bandwidth unfiltered pulses are shown exhibiting the characteristic bipolar structure, while
in the right panel an approximate response of the ANITA III antennas is applied. The pulses are qualitatively similar to those detected in ANITA III for events that are compatible with being produced by atmosphere-skimming showers, i.e. the above-horizon events b) and c) in figure 3 of \cite{ANITAIII}.

%%%%%%%%%%%%%%%%%%%%%%%%%%%%%%%%
%%%%%%%%%%%%%%%%%%%%%%%%%%%%%%%%%%%%%%%%%%%%

%%%%%%%%%%%%%%%%%%%%%%%%%%%%%%%%%%%%%%%%%
\subsection{Lateral distribution of the radio signal}

\begin{figure}[htb]
    \centering
    \includegraphics[width = .9\textwidth]{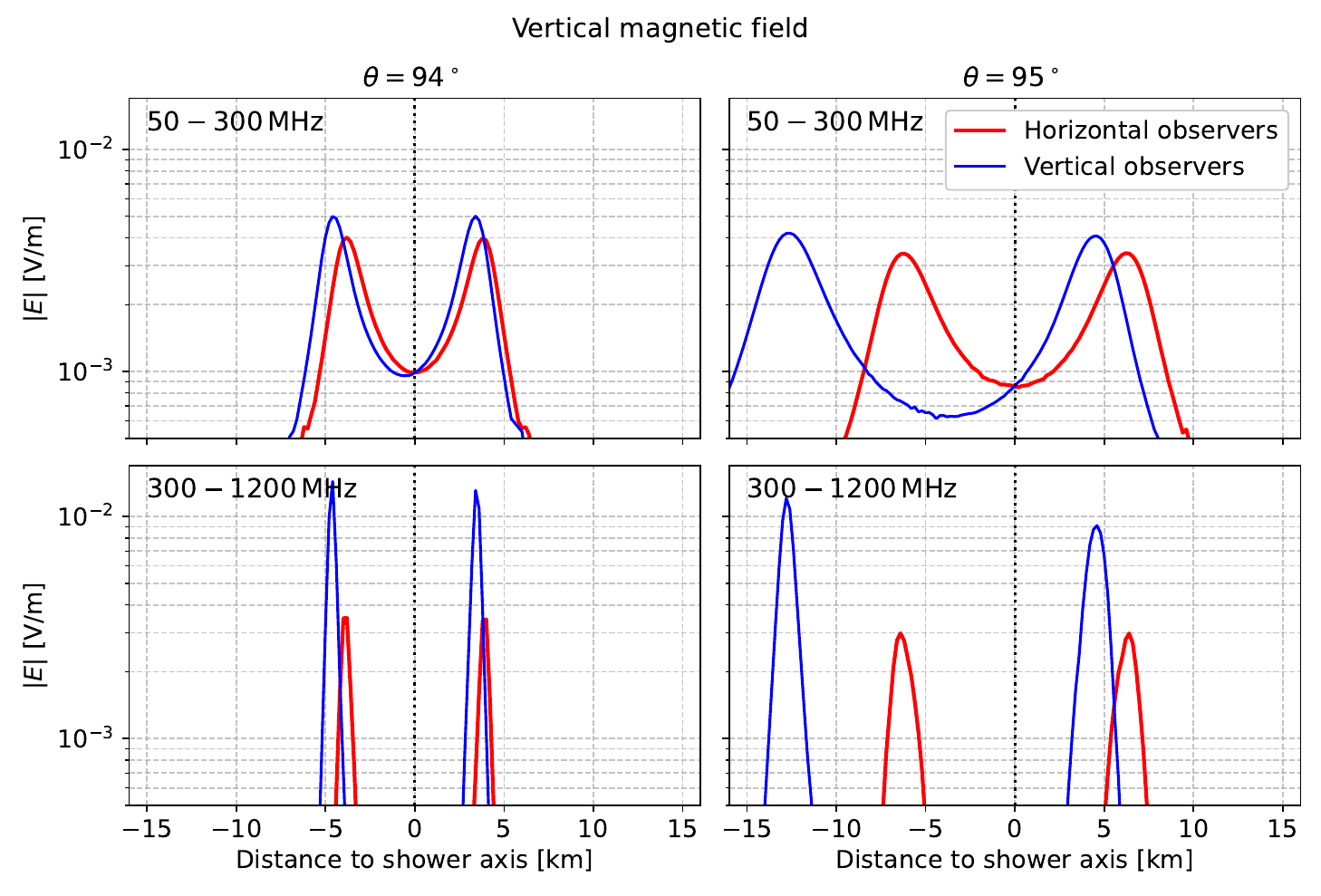}
    \caption{Lateral distribution (LDF) of the maximum electric field modulus around shower axis, obtained in proton-induced showers of energy $10^{19.5}\unit{eV}$, propagating under the effect of a vertical magnetic field of  $50\unit{\mu T}$ as sketched in the top panel of figure \ref{fig:RASPASSAntennas} and passing at $h=36\unit{km}$ a.s.l. The LDF is shown both in antennas placed in the directions \textit{horizontal} or parallel to ground (red lines) and \textit{vertical} or perpendicular to ground (blue lines), as indicated in figure \ref{fig:RASPASSAntennas}, for showers of zenith angle $\theta = 94^\circ$ (left panels) and $95^\circ$ (right panels). The LDF is shown in two different frequency bands, namely  $50-300\unit{MHz}$ (upper panels) and $300-1200\unit{MHz}$ (lower panels).}
    \label{fig:LDFBVer}
\end{figure}

\begin{figure}[htb]
    \centering
    \includegraphics[width = .9\textwidth]{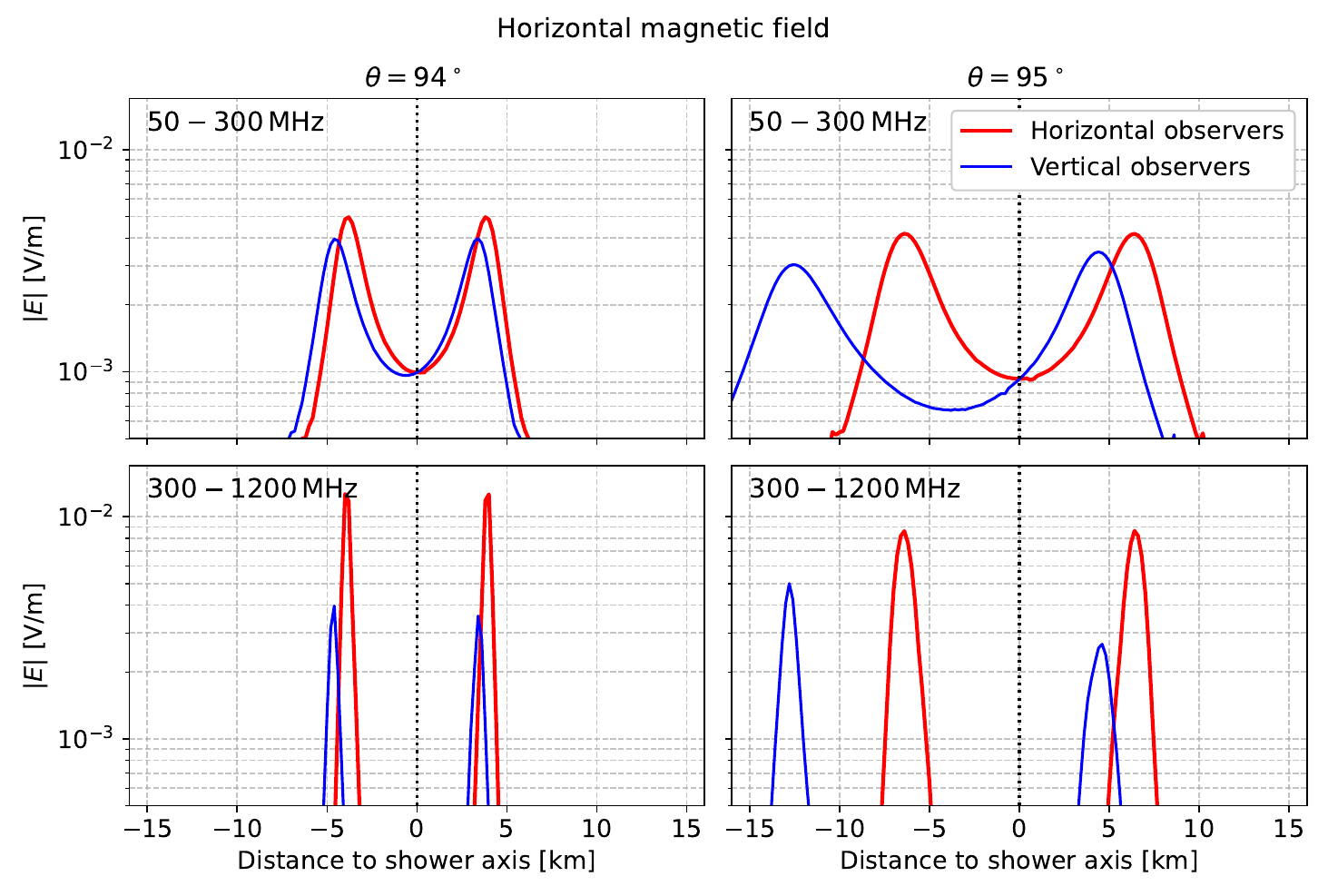}
    \caption{Same as figure \ref{fig:LDFBVer}, considering showers of the same geometries and primary energy propagating under a horizontal (parallel to ground) magnetic field of $50\unit{\mu T}$ as sketched in the bottom panel of figure \ref{fig:RASPASSAntennas}.}
    \label{fig:LDFBHor}
\end{figure}

\begin{figure}[htb]
    \centering
    \includegraphics[width=.75\linewidth]{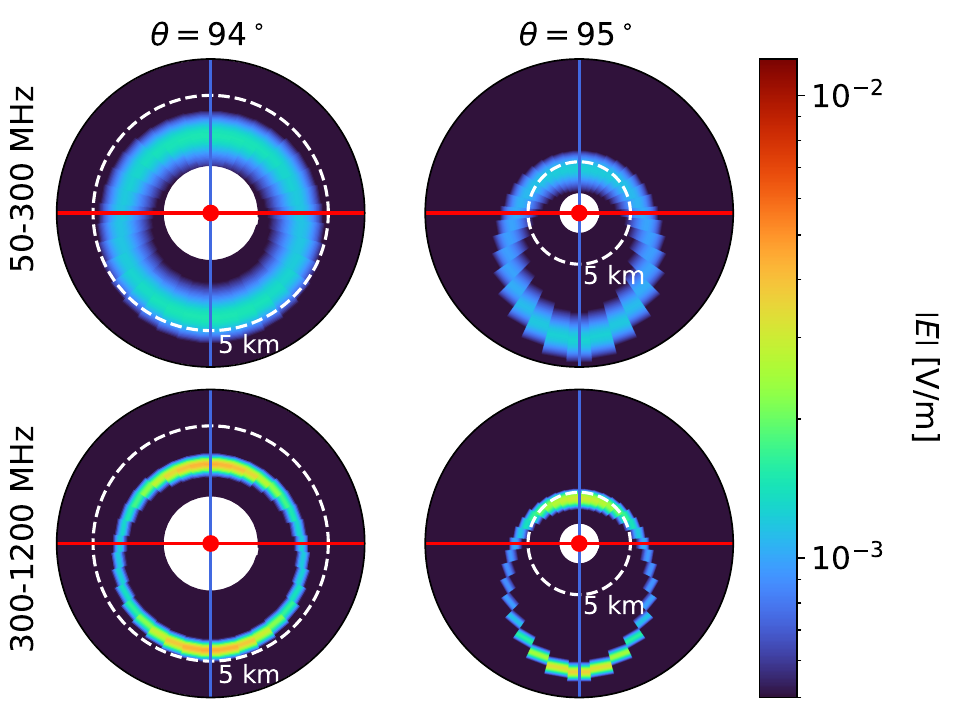}
    \caption{Two-dimensional distribution of the maximum electric field modulus (color scale) in the plane 
    perpendicular to the shower axis. The results are shown for proton-induced showers with energy $E_0=10^{19}\unit{eV}$, at zenith angles $\theta = 94^\circ$ (left panels) and $95^\circ$ (right panels), propagating along the north-south direction in a magnetic field of inclination $I=-70^\circ$ and passing at a height $h=36\unit{km}$. The peak electric field is shown in two frequency ranges, $50-300\unit{MHz}$ in the upper panels and $300-1200\unit{MHz}$ in the lower panels. The horizontal red and vertical blue lines are approximately equivalent to the horizontal and vertical observer directions in figure \ref{fig:RASPASSAntennas}, while a reference white dashed line indicates the positions that lie $5\unit{km}$ away from the shower axis (red dot) in each case, highlighting the different distance scales. The white zone around shower core corresponds to regions where no antennas were placed in the simulations. The refractive and coherence asymmetries are apparent (see text for explanations). The discontinuities in the electric field distributions are an artifact of the representation in polar coordinates.}
    \label{fig:LDF2D}
\end{figure}

\begin{figure}[htb]
    \centering
    \includegraphics[width=.75\linewidth]{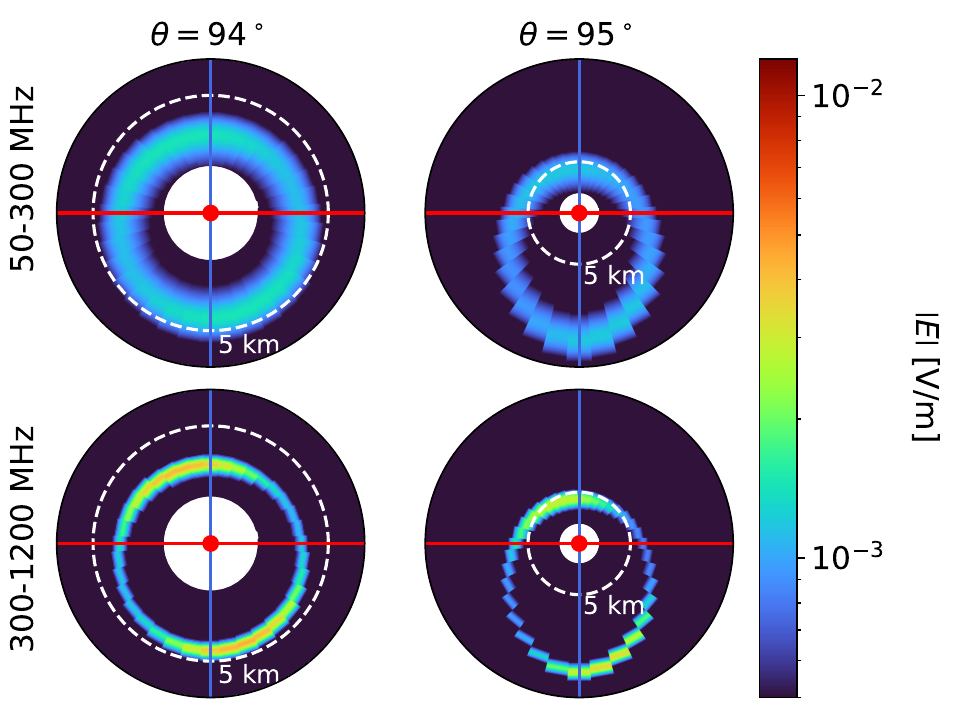}
    \caption{Same as figure \ref{fig:LDF2D} for air showers propagating along the west-east direction under the same magnetic field of inclination $I=-70^\circ$
    }
    \label{fig:LDF2D_BInc}
\end{figure}

In figures \ref{fig:LDFBVer} and \ref{fig:LDFBHor} we show the simulated lateral distribution (LDF) of the total electric field amplitude along the horizontal and vertical directions w.r.t. the ground plane (see figure \ref{fig:RASPASSAntennas}), after filtering in the $50-300\unit{MHz}$ and $300-1200\unit{MHz}$ frequency bands, respectively. The chosen frequency ranges 
correspond approximately to the proposed operational bands of the Low-Frequency (LF) and Main (MI) instruments in PUEO \cite{PUEO}, the latter being similar to the band used in the ANITA flights ($200-1200\unit{MHz}$). In addition, the LDFs were obtained for two different magnetic field orientations: \textit{vertical} or perpendicular to ground with an inclination $I=-90^\circ$ (left panels) and \textit{horizontal} or parallel to the ground plane with an inclination $I=0^\circ$ (right panels). 
These configurations serve as an example of the typical geomagnetic field either close to the magnetic poles or at more equatorial latitudes, respectively.
In figures \ref{fig:LDF2D} and \ref{fig:LDF2D_BInc} we also show the two-dimensional lateral distribution function in a plane perpendicular to shower axis, for the case of proton air showers propagating with different orientations, in a magnetic field of inclination $I=-70^\circ$ similar to that at the geographic South Pole, where balloon-borne experiments are commonly flown.

%----------------------------------------------------------------------
The observer positions where the coherence of the signal is maximal form a ring-like region around shower axis as shown in figures \ref{fig:LDF2D} and \ref{fig:LDF2D_BInc}, which explains the appearance of two peaks in the LDFs shown in figures \ref{fig:LDFBVer} and \ref{fig:LDFBHor} for observers located along the vertical or horizontal directions (see figure \ref{fig:RASPASSAntennas}). The difference in the position, amplitude and width of the peaks in the respective LDFs, as well as in the size of the ring-like region and amplitude of the electric field in figures \ref{fig:LDF2D} and \ref{fig:LDF2D_BInc}, is produced by the slight difference in geometry of the showers, $\theta=94^\circ$ in the left and $\theta=95^\circ$ in the right panels, that leads to a significant change in propagation conditions.

In general terms, the larger the zenith angle, showers travel in a denser atmosphere where the Cherenkov angle, that serves as a good estimation of $\psi_{\rm max}$, is larger, and they also develop farther from the detector\footnote{$X_{\rm max}$ for $\theta=94^\circ$ ($95^\circ$) occurs at a distance $\simeq 620\unit{km}$ ($\simeq 886\unit{km}$) from the detector, corresponding to altitudes $\simeq 22.7$ ($\simeq 20\unit{km}$) above sea level.}. Due to these two effects, the size of the ring-like region where the coherence is maximal is larger for showers with $\theta=95^\circ$ when compared to $\theta=94^\circ$, as can be seen in figures \ref{fig:LDFBVer} to  \ref{fig:LDF2D_BInc}. 
Also, due to the larger distance from $X_\mathrm{max}$ to the detector for $\theta=95^\circ$ the signal gets more attenuated, an effect contributing to the amplitude of the electric field at the ring, with peaks being smaller than for $\theta=94^\circ$.
The rapid falloff of the electric field amplitude in the $300-1200\unit{MHz}$ band as the observer moves away from $\psi_{\rm max}$ (bottom panels of figures \ref{fig:LDFBVer} to \ref{fig:LDF2D_BInc}) compared to the $50-300\unit{MHz}$ band (top panels) is due to the destructive interference of emissions from different regions in the shower development, with the high-frequency components adding up less coherently for equivalent deviations from $\psi_{\rm max}$ \cite{Zas:1991jv}.

%%%%%%%%%%%%%%%%%%%%%%%%%%%%%%%%%%%%%%%%%
\subsubsection{Refractive asymmetry}
\label{sec:refractive_asym}

An asymmetry in the lateral distributions of electric field amplitudes with respect to the observer position around shower axis becomes evident when comparing the LDFs along the horizontal and vertical directions. The LDFs along the direction parallel to ground (\textit{Horizontal} axis in red in figure \ref{fig:RASPASSAntennas} and red lines in figures \ref{fig:LDFBVer} and \ref{fig:LDFBHor}) are almost symmetric around the shower axis, while 
the LDFs in the direction perpendicular to ground (\textit{Vertical} axis in blue in figure \ref{fig:RASPASSAntennas} and blue lines in figures \ref{fig:LDFBVer} and \ref{fig:LDFBHor}) appear displaced towards lower altitudes, corresponding to negative values of the lateral distance. This asymmetry is also visible in the two-dimensional LDFs shown in figures \ref{fig:LDF2D} and \ref{fig:LDF2D_BInc}.

The displacement has its roots on the propagation of signals across the changing refractive index following the density gradient in the atmosphere \cite{Refractive_displacement}. Since the propagation times of the signals from the air shower to observers depend on the refractive index profile through which the signal travels, the gradient of index of refraction can alter the relative time delays between emissions, and thus the coherence of the observed pulses, depending on the position of the observer.

For two observers placed symmetrically around the shower axis along the \textit{horizontal} direction (red axis in figure \ref{fig:RASPASSAntennas}), the distance from the shower axis to the observers and the refractive index profile traversed by the signals are identical.  
In contrast, for two observers placed at the same distance from the shower axis along the \textit{vertical} direction (blue axis in figure \ref{fig:RASPASSAntennas}), the optical paths are not the same due to the dependence of the atmospheric refractive index with altitude. In this case, rays propagating from the shower to observers placed at higher altitude traverse regions of lower density and refractivity than those propagating towards observers at lower altitudes. By means of a simple toy Monte Carlo approximating the shower as a one-dimensional distribution, and for the geometries considered in this work, we have explicitly checked that the propagation along different directions through different profiles of refractive index affects the relative time delays between emissions at the observers, producing the refractive asymmetry predicted by the full RASPASS simulation.

Inspecting the blue lines in figures \ref{fig:LDFBVer} and \ref{fig:LDFBHor} and for the observers placed along the blue line in figures \ref{fig:LDF2D} and \ref{fig:LDF2D_BInc}, it can be seen that the displacement of the peaks of the LDF in the vertical direction is enhanced as $\theta$ increases, since the difference in the optical paths between the shower and observers above and below the shower axis increases with zenith angle. The magnitude of the displacement and its dependence with $\theta$ is consistent with the positions where the accumulated time delays between emissions predicted by the one dimensional Monte Carlo approach are minimal, and thus where the signal would be expected to reach a maximal degree of coherence. The predicted positions correspond approximately to the positions of the peaks of the LDFs in figures \ref{fig:LDFBVer} and \ref{fig:LDFBHor} obtained with the full RASPASS simulations.

The asymmetry due to the changing refractive index is only weakly dependent on the observation frequency band since the atmosphere is assumed to be non-dispersive in the frequency ranges considered \cite{Refractive_displacement}. This can be checked by comparing the position of the peaks of the LDF in the top and bottom panels in figures \ref{fig:LDFBVer} and \ref{fig:LDFBHor}, where the signal is filtered in different frequency bands.

In ZHAireS, the signal propagation between emission and observer is assumed to follow a straight path, neglecting the bending of rays in a inhomogeneous atmosphere. Using the one-dimensional Monte Carlo approach mentioned before, the difference between propagation along curved or straight paths was found to produce a negligible difference in the spread of the distribution of arrival times of individual emissions, even for $\theta = 95^\circ$ for which the gradients in density and refractivity are larger. For this reason, we do not expect that the approximation of propagation along straight tracks would significantly affect the refractive asymmetry observed in the simulations.

%%%%%%%%%%%%%%%%%%%%%%%%%%%%%%%%%%%%%%%%%
\subsubsection{Coherence asymmetry}
\label{sec:coherence_asym}

The important role played by the Earth's magnetic field in the development of AS showers induces a second asymmetry in the lateral distribution of the radio emission. 
As shown in figures \ref{fig:LDFBVer} and \ref{fig:LDFBHor}, for a given zenith angle and frequency band, the magnitude of the simulated electric field at the peaks of the LDF is different depending on whether the observer is placed along the horizontal or vertical to the ground (figure \ref{fig:RASPASSAntennas}). This asymmetry is also dependent on the orientation of the magnetic field as can be seen comparing figures \ref{fig:LDFBVer} and \ref{fig:LDFBHor}. The maximum electric field is registered by observers along the vertical direction for a magnetic field perpendicular to the ground plane (figure \ref{fig:LDFBVer}), or along the horizontal direction for the case of a magnetic field  parallel to the ground plane (figure \ref{fig:LDFBHor}).

The origin of this asymmetry lies in the spread of the shower front preferentially in the direction $\vect{v}\cross\vect{B}$, caused by the deflection of particles in the Earth's magnetic field \cite{Tueros_Showers_JCAP, Chiche:2024yos}. The large distances and low atmospheric density over which atmosphere-skimming cascades develop allow for the geomagnetic field to produce a significant charge separation. The net effect is a flattening of the shower front along the direction of the Lorentz force as sketched in figure \ref{fig:RASPASSAntennas} and exemplified in figure \ref{fig:LongLatDev_Bfield}. The growth of the lateral dimension of the shower along the direction $\vect{v}\cross\vect{B}$ when compared to the direction parallel to $\vect{B}$ has the effect of increasing, by purely geometrical reasons and depending on the position of the observer, the time delays between contributions of different parts of the shower to the total signal.

For observers placed in the plane containing the $\vect{v}\cross\vect{B}$ vector the time delays between different parts of the shower front are typically larger than those seen by observers outside that plane. 
Increasing the time delays limits the frequencies at which the Fourier components of the electric field add up coherently to contribute to the total emission, typically reducing the higher frequency content of signals and their amplitude. 

This interpretation allows to understand the behaviour seen in figures \ref{fig:LDFBVer} and \ref{fig:LDFBHor}, and in the two-dimensional distributions of figures \ref{fig:LDF2D} and \ref{fig:LDF2D_BInc}. If the magnetic field is oriented in the direction perpendicular to ground (figure \ref{fig:LDFBVer}), the shower front spreads along a direction almost \textit{horizontal} (parallel to ground) as sketched in the top panel of figure \ref{fig:RASPASSAntennas}. This in turn increases the time delays for observers placed along the \textit{horizontal} direction (red line in figure \ref{fig:RASPASSAntennas}) compared to observers placed along the \textit{vertical} direction. This effect leads to larger signals at the peaks of the radio LDF for the vertical observers. The effect is reversed when the magnetic field is oriented in the horizontal direction, as can be seen in figure \ref{fig:LDFBHor} where the largest contribution to the peak amplitude is registered by observers in the horizontal direction. In the case of showers propagating under the effect of a magnetic field with an intermediate inclination between horizontal and vertical, the positions that receive the larger signal depend on the incoming direction of the air shower relative to the magnetic north. 
An example of this effect is shown in figures \ref{fig:LDF2D} and \ref{fig:LDF2D_BInc} for the case of a magnetic field with inclination $I=-70^\circ$. In this case, showers propagating from the magnetic north towards south (figure \ref{fig:LDF2D}) are spread in a direction parallel to ground (represented by the red line in figure \ref{fig:LDF2D}). As a consequence, the strongest electric fields appear along the vertical direction, exactly above and below the shower axis. On the other hand, when the shower propagates from magnetic west towards east (figure \ref{fig:LDF2D_BInc}), the $\vect{v}\times\vect{B}$ direction has a deviation of $20^\circ$ relative to the horizontal as a consequence of the inclination of the magnetic field. Thus, the largest signals appear in this case \textit{rotated} by the same angle with respect to the vertical direction. The fact that the general shape of the ring-like region does not change with the inclination of the magnetic field indicates that the \textit{coherence asymmetry} and the \textit{refractive asymmetry} are independent phenomena.

Unlike the refractive asymmetry of the LDF, the coherence asymmetry is naturally dependent on the frequency band of observation, having its main impact at the highest frequencies. This can be checked by comparing the relative difference in the peak electric field between the horizontal and vertical observers in the two frequency bands shown in figures \ref{fig:LDFBVer} and \ref{fig:LDFBHor}. For example, for the case of a shower with $\theta = 94^\circ$ propagating under the effect of a vertical magnetic field (left panels in figure \ref{fig:LDFBVer}), the relative difference between the peak of the LDF in the vertical and horizontal observation direction\footnote{Defined as  $\displaystyle \left(E_V - E_H\right)/\left(E_H+E_V\right)$ with $E_H$ ($E_V$) the maximum amplitude of the electric field in the horizontal (vertical) direction.} is $\sim10\%$ in the $50-300\unit{MHz}$ band increasing to $\sim60\%$ in the $300-1200 \unit{MHz}$ frequency band. 

The effect of the coherence asymmetry also depends on the shower zenith angle. As $\theta$ decreases, showers propagate at higher altitudes where they need to travel increasingly larger distances to develop due to the lower atmospheric density profile they traverse. This gives more time for the magnetic field to spread the shower front in a lower density medium where the magnetic deflection of the particle trajectories is stronger.
As a consequence, the coherence asymmetry is expected to increase in showers with smaller values of $\theta$ closer to $\theta=90^\circ$.
A comparison between the peak electric field along the horizontal or vertical direction in the $300-1200\unit{MHz}$ frequency band (figure \ref{fig:LDFBHor}) reveals that the asymmetry increases from $\sim \pm\,50\%$ for $\theta = 95^\circ$ to $\sim \pm \,60\%$ for $\theta = 94^\circ$.

\begin{figure}[tb]
   \centering
    \includegraphics[width=.75\linewidth]{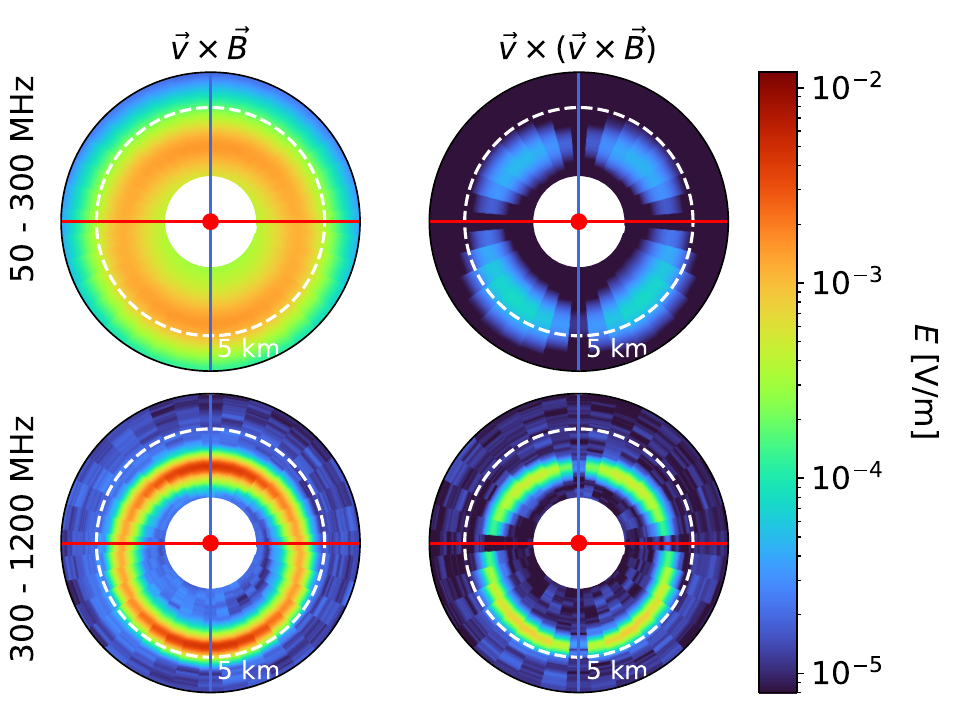}
    \caption{Two-dimensional distribution of the maximum electric field (color scale) in the plane perpendicular to the shower axis, distinguishing the polarizations parallel to the $\vect{v}\times\vect{B}$ (left panels) or $\vect{v}\times\left(\vect{v}\times\vect{B}\right)$ direction (right panels); and in two different frequency ranges, $50-300\unit{MHz}$ in the upper panels and $300-1200\unit{MHz}$ in the lower panels. The results are shown for the case of a proton-induced shower with energy $E_0=10^{19}\unit{eV}$ and $\theta = 94^\circ$ propagating along the north-south direction in a magnetic field of inclination $I=-70^\circ$ and passing at a height $h=36\unit{km}$. The horizontal red and vertical blue lines are approximately equivalent to the horizontal and vertical directions in figure \ref{fig:RASPASSAntennas}, while a reference white dashed line indicates the positions that lie $5\unit{km}$ away from the shower axis (red dot) in each case. The white region around shower core corresponds to observer positions where no antennas were placed in the simulations.}
    \label{fig:LDF2D_polzs}
\end{figure}

A loss of coherence of the radio emission was also predicted with ZHAireS simulations in the case of inclined downward-going showers compared to more vertical ones. This was mainly attributed to the wider lateral spread of inclined showers due to the magnetic deflection in the lower density layers of the atmosphere \cite{Ammerman-Yebra:2023rhr,Chiche:2024yos}. For AS showers, the coherence loss with respect to downward-going vertical showers has been studied with RASPASS simulations, being even larger than in inclined downward-going showers and following the same trend with atmospheric density as that obtained in \cite{Chiche:2024yos}.

Finally, in figure \ref{fig:LDF2D_polzs} we show, for the same proton shower with $\theta = 94^\circ$ represented in figure \ref{fig:LDF2D}, the distribution around shower axis of the maximum amplitude of the electric field along the two polarizations orthogonal to the shower incoming direction,  perpendicular  ($\vect{v}\times\vect{B}$, left panels), or parallel to the magnetic field ($\vect{v}\times\left(\vect{v}\times\vect{B}\right)$, right panels); and in two different frequency ranges. As expected, the electric field is predominantly polarized along the $\vect{v}\times\vect{B}$ direction, demonstrating the dominance of the geomagnetic mechanism as the main origin of the emitted radiation in the particular case of AS showers; while the magnitude of the electric field polarized along the $\vect{v}\times\left(\vect{v}\times\vect{B}\right)$ direction is orders of magnitude smaller. The lateral distributions for both polarizations feature the same refractive displacement, as expected. Also, the maximum amplitude of the $\vect{v}\times\left(\vect{v}\times\vect{B}\right)$ polarization is distributed around axis in a four-lobed pattern similar to that reported in \cite{Chiche:2024yos} for the case of inclined downward-going air showers. 

%%%%%%%%%%%%%%%%%%%%%%%%%%%%%%%%%%%%%%%%%
\subsection{Frequency spectrum of the radio signal}

\begin{figure}
    \centering
    \includegraphics[width = \textwidth]{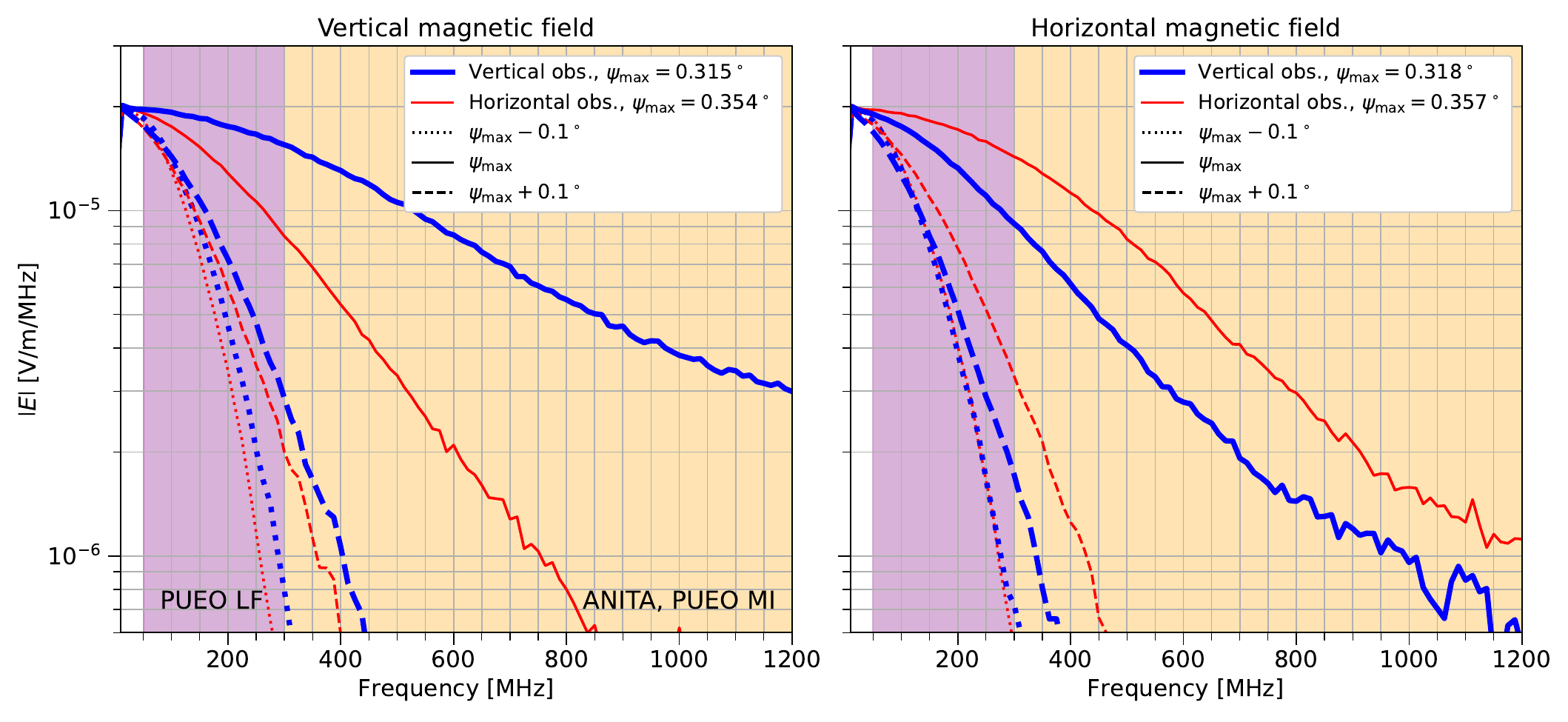}
    \caption{Frequency spectrum of the simulated electric field obtained in proton-induced showers of primary energy $10^{19.5}\unit{eV}$ and zenith angle $\theta = 94^\circ$, passing at $36\unit{km}$ a.s.l. under the effect of a magnetic field of $50\unit{\mu T}$ pointing in the vertical (left panel) or horizontal direction (right panel). The electric fields are shown for observers both along the directions horizontal or parallel to ground (red thin lines) and vertical (blue thick lines), either at the angle $\psi_{\rm max}$ where the coherence of the signal is maximal (solid lines), or $0.1^\circ$ inwards-outwards of this position (dotted and dashed lines respectively). The two shaded regions represent the frequency range of both the PUEO Low-Frequency (LF, $50-300\unit{MHz}$) and Main (MI, $300-1200\unit{MHz}$) Instruments, the latter being similar to the frequency range of ANITA ($200-1200\unit{MHz}$). The different values of $\psi_{\rm max}$ along the two directions is a consequence of the refractive asymmetry.}
    \label{fig:Spectra}
\end{figure}

Further insight into the characteristics of radio emission in AS showers can be gained by looking at the frequency spectra of the simulated signals. In figure \ref{fig:Spectra} we show the frequency spectrum in a proton shower with $\theta = 94^\circ$ propagating either in a vertical or horizontal magnetic field of intensity $50\unit{\mu T}$. The frequency spectrum of the electric field amplitude is shown at the viewing angle $\psi_{\rm max}$ where the coherence of the signal is maximal (where the LDFs reach their maximum values in figures \ref{fig:LDFBVer} and \ref{fig:LDFBHor}), and at $\psi_{\rm max} \pm \,0.1^\circ$.

%------------------------------------------------------------------------
%-----------------------------------------------------------------------------------------

The spectra are shown both for observers placed along the horizontal direction (red thin lines) or seeing the shower from above along the vertical direction (blue thick lines), following the same nomenclature as in figure \ref{fig:RASPASSAntennas}. Observers along the horizontal direction are placed at an altitude of $36\unit{km}$ a.s.l., while the observers seeing the shower from above at angles $\psi_{\rm max}-0.1^\circ$, $\psi_{\rm max}$ and $\psi_{\rm max}+0.1^\circ$ are located at altitudes above sea level of $\sim 38.4\unit{km}$, $39.4\unit{km}$ and $40.4\unit{km}$ respectively.

All the spectra shown in figure \ref{fig:Spectra} feature an approximately exponential fall-off of the field amplitude with frequency. This behaviour is similar to what is observed in the radiation emitted by downward-going air showers and reflected off the ice cap of Antarctica \cite{Reflex, ANITAEnergyFlux}. The exponential fall-off seen in ZHAireS-RASPASS simulations is also consistent with the direct cosmic ray signals detected in the ANITA flights \cite{ANITAEnergyFlux}. 

The effect of the coherence asymmetry is clearly visible when comparing the spectra at the maximum coherence angle $\psi_{\rm max}$ between observers placed along the horizontal or vertical direction. For the vertical magnetic field configuration in the left panel of figure \ref{fig:Spectra}, the shower is \textit{flattened} in a plane almost parallel to ground (upper panel in figure \ref{fig:RASPASSAntennas}), affecting the high frequency content of the pulses seen by observers along the horizontal direction. When the magnetic field is oriented parallel to ground and the shower front spreads in a vertical plane (lower panel in figure \ref{fig:RASPASSAntennas}), the effect is reversed as can be seen in the right panel of figure \ref{fig:Spectra}. The resulting effect is a more rapid fall-off with frequency of the electric field amplitude for observers inside the plane 
containing the $\vect{v}\cross\vect{B}$ vector, due to the lateral spread of the shower front.

For observers located at off-axis angles $\psi_{\rm max}\pm0.1^\circ$, the increased time delays accumulated between different parts of the shower development quickly remove high-frequency content from the signal \cite{Zas:1991jv}, which in turn becomes less sensitive to the effect of the coherence asymmetry. The reduction in the high-frequency content of the pulses as the observer moves away from the angle $\psi_{\rm max}$ is much more pronounced than in the case of signals produced by particle cascades with downward-going trajectories, since the length scales of atmosphere-skimming showers, that determine the magnitude of the time delays, are significantly larger than those of downward-going air showers \cite{Tueros_Showers_JCAP}.

%%%%%%%%%%%%%%%%%%%%%%%%%%%%%%%%%%%%%%%%%%%%%%%%%%
\section{Implications for radio detection in high-altitude experiments}
\label{sec:implications}

In this section we explore the implications of the properties of radio emission in AS showers for their detection in high-altitude balloon-borne experiments. We concentrate on the energy reconstruction of AS showers and on the expected effective area of balloon-borne experiments.

%%%%%%%%%%%%%%%%%%%%%%%%%%%%%%%%%%%%%%%%%%%%%%%%%%
\subsection{Systematic uncertainties in shower energy determination}

A method for shower energy reconstruction was introduced in \cite{ANITAEnergyFlux} in the context of the ANITA experiment, for the case of cosmic-ray induced showers whose radio emission reflected on the ground before reaching the detector. For the first time, a measurement of the flux of cosmic rays was performed using the radio technique, in agreement with other experimental observations \cite{ANITAEnergyFlux}. Simulations of radio emission performed with ZHAireS for these reflected events revealed that the slope of the frequency spectrum in the ANITA frequency range (200 MHz - 1200 MHz) is sensitive to the off-axis angle under which the shower maximum is observed \cite{Reflex}. This makes it possible to obtain the viewing angle from measurements of the frequency spectrum of the radio emission. Once this is known, the maximally-coherent pulse amplitude (at off-axis $\psi=\psi_{\rm max}$), that scales linearly with the primary energy, can be inferred providing a measurement of the shower energy \cite{ANITAEnergyFlux}.

\begin{figure}
    \centering
    \includegraphics[width = \textwidth]{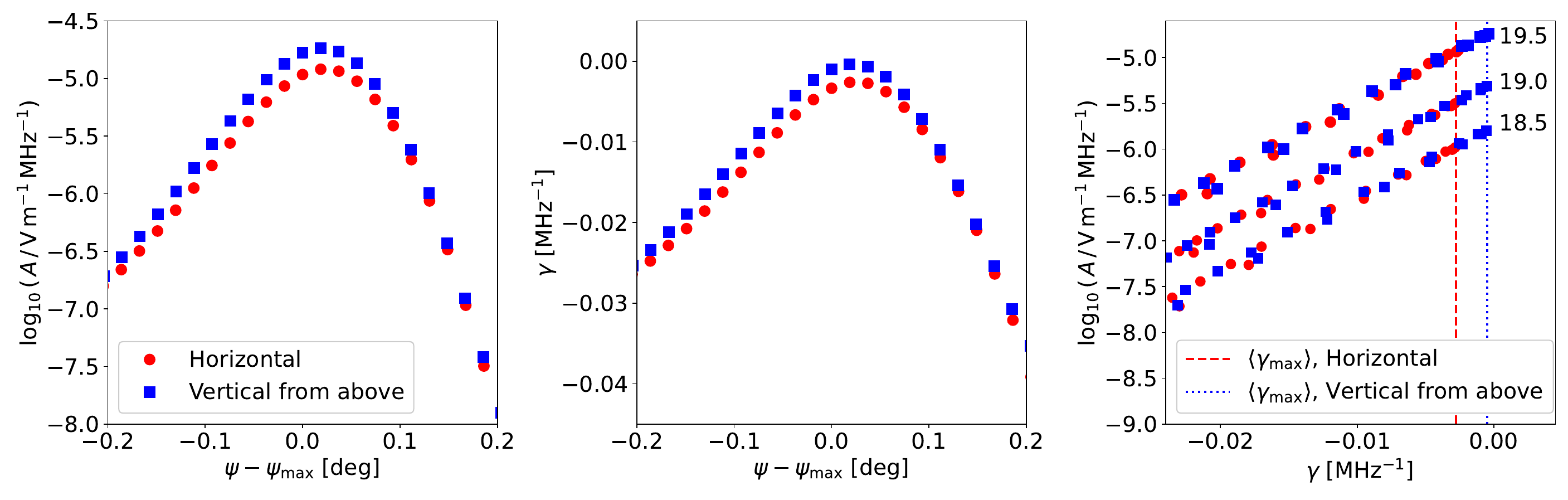}
    \caption{Results of fits of the frequency spectrum of the simulated electric field within the $100-250\,\mathrm{MHz}$ frequency band to equation \eqref{eq:ExpFit}. The frequency spectrum was obtained for proton showers with zenith angle $\theta = 94^\circ$ passing at an altitude $h=36\unit{km}$ a.s.l., under the effect of a magnetic field of $50\unit{\mu T}$ perpendicular to ground. Left panel: Amplitude parameter $\log_{10}A$ as a function of the off-axis angle w.r.t. the maximum coherence angle $\psi-\psi_{\rm max}$, in a shower of primary energy $10^{19.5}\unit{eV}$, both for observers seeing the shower along the vertical (from above) and horizontal directions. Center panel: Same as left panel for the slope parameter $\gamma$. Right panel: Correlation between the fitted $\log_{10}A$ and $\gamma$ for the same observers as in the other panels, for showers with three different primary energies, labeling the curves with the value of  $\log_{10}{E_0\,\left(\mathrm{eV}\right)}$. The dashed red (dotted blue) line indicates the average maximal spectral slope ($\gamma_{\rm max}$) for observers seeing the shower along the parallel to ground (vertical) direction. The error bars associated to the exponential fit are smaller than markers in all panels. Shower to shower fluctuations were not considered.}
    \label{fig:Spectral_fits}
\end{figure}

To study the applicability of a similar energy reconstruction methodology to AS showers, we fitted the simulated electric fields in the $100-250\unit{MHz}$ frequency band using an exponential function similar to that employed in \cite{ANITAEnergyFlux},
\begin{equation}
    \left|E\left(f\right)\right| = A\,\exp\left[\gamma\left(f-250\unit{MHz}\right)\right]\,.
    \label{eq:ExpFit}
\end{equation}
The results of the fit for $A$ (amplitude at $250\unit{MHz}$), and $\gamma$ (spectral slope), are shown in the left and center panels of figure \ref{fig:Spectral_fits} as a function of the observation angle. As expected, the amplitude $A$ and slope $\gamma$ are largest at $\psi = \psi_{\rm max}$ decreasing as the observer moves away from it. Also at $\psi = \psi_{\rm max}$, there is a clear difference in the amplitude $A$ and spectral slope $\gamma$ for observers placed along the direction of $\vect{v}\cross\vect{B}$ (\textit{Horizontal} in figure \ref{fig:Spectral_fits}) or along the perpendicular to it (\textit{Vertical from above} in figure \ref{fig:Spectral_fits}). This difference decreases for observation angles away from $\psi_{\rm max}$. The origin of this difference is the coherence asymmetry explained in Section \ref{sec:coherence_asym}.

%-----------------------------------------------------------------------------------------
%--------------------------------------------------------------------------------------------

To study the impact that the position of the observer has on the determination of shower energy, in the right panel of figure \ref{fig:Spectral_fits}, we plot the correlation between $A$ and $\gamma$ for observers along the vertical and horizontal direction, for different primary energies. The almost linear correlation between $\log_{10}A$ and $\gamma$ serves as a calibration of the energy of the event \cite{ANITAEnergyFlux}. Once $A$ and $\gamma$ are estimated from the measured signal the energy can be obtained from the corresponding calibration line. 
Although the calibration lines obtained with observing positions either along the vertical (blue squares) or horizontal (red dots) directions are rather similar, a significant difference in the maximum values of the spectral amplitude and slope appears depending on the observer position, as a consequence of the coherence asymmetry. The maximum slope $\gamma_{\rm max}$ is indicated by the dashed red and blue dotted lines in the right panel of figure \ref{fig:Spectral_fits}, as also seen in figure \ref{fig:Spectra}. Since the determination of the shower energy relies on the estimation of the amplitude of the maximally-coherent pulse, this would lead to an uncertainty in the determination of shower energy associated to the position of the observer, that could reach up to $\sim 20\%$ in the $100-250\unit{MHz}$ frequency range where the fits were performed.

These results show how small changes in the geometry of the shower, relative to the position of the detector and the local magnetic field, can significantly affect the properties of radio signal and its normalization, as well as the lateral distribution of the electric field. This calls for a careful selection of the simulated events needed to estimate shower energy in order to properly take into account the effects of the refractive and coherence asymmetries, by considering shower geometries that are compatible with the direction of the observed pulses.

%%%%%%%%%%%%%%%%%%%%%%%%%%%%%%%%%%%%%%%%%%%%%%%%%%%%%%%%%%%%%%%%
\subsection{Effective area of balloon-borne experiments to AS showers.}
\label{sec:Area}

The geometry of atmosphere-skimming showers introduces complex effects in the calculation of the effective area for high-altitude experiments using the radio technique. As discussed in the previous section, the refractive and coherence asymmetries affect differently the properties of the signal depending on the position of the detector and its frequency band of operation, the arrival direction of the primary particle and the orientation of the magnetic field. The complex interplay between these factors calls for detailed simulations to study how the effective area depends on the design, characteristics and location of the detector.
 
We illustrate these dependencies using the reported directions of the two atmosphere-skimming events detected during the ANITA IV flight, event “A" with ID 9734523 and event “B" with ID 51293223 \cite{ANITAIV}.
For each of the two events, we performed a set of simulations of AS showers with either proton or iron as primary particle, with different energies and impact parameter $b$ (see figure \ref{fig:RASPASSAntennas}). The simulations were carried out with the aim of showing the effects of the refractive and coherence asymmetries on the effective area, rather than attempting to reconstruct the properties of each event. 
The zenith and azimuth angles of the simulated showers ($\theta$, $\phi$) were fixed to the direction of the received pulses for events A ($\theta_{\rm A}=95.64^\circ$, $\phi_{\rm A}=2.01^\circ$) and B ($\theta_{\rm B}=95.38^\circ$, $\phi_{\rm B}=306.45^\circ$) as reported in \cite{ANITAIV}. The azimuth was assumed relative to the magnetic north.  
For each of the two directions simulated, the energy $E_0$ and altitude of the shower axis $h$ was varied (and thus the impact parameter $b$, figure \ref{fig:RASPASSAntennas}).
Realistic atmospheric density and refractivity profiles were set according to the date of the events using the GDAS database \cite{GDAS} in the range of altitudes where air showers with these geometries develop. Also, the magnetic field was assumed to be the one at the South Pole at the date of the events, which was obtained using the IGRF13 model \cite{IGRF13}.

For every simulated shower with parameters $(E_0,\theta,\phi,h)$, the amplitude of the electric field in the $300-1000\unit{MHz}$ frequency band was found at the position of a detector hovering at the altitudes $h_{\rm det, A}=39.25\unit{km}$ and  $h_{\rm det,B}=37.53\unit{km}$ reported in \cite{ANITAIV}. We assumed a typical threshold value for triggering of $325\unit{\mu V/m}$ at the peak of the field \cite{ANITAIVTrigger} in the frequency band.
In figure \ref{fig:ANITAIVEvents}, we plot the positions of the shower axes such that a trigger is produced, in the plane $X-Y$ containing the detector (located at $(X, Y)=(0,0)$) and perpendicular to the projection of the shower axis on the ground plane. This is done for the incoming directions of events A (top panels) and B (bottom panels). The colour scale indicates the magnitude of the peak electric field at the detector. As an example, a $E_0=10^{19}\unit{eV}$ proton shower with the incoming direction of event A (upper middle panel in figure \ref{fig:ANITAIVEvents}) whose axis passes at $Y=-3\unit{km}$ (below the payload, as indicated with a green dot) would generate an electric field of $\sim1.6\unit{mV/m}$ at the detector, enough to produce a trigger. However, if the shower axis were to pass at $Y=+3\unit{km}$ (above the detector as indicated with a red dot), the signal would be too small to trigger the detector. The locations of the axes of the showers whose signals trigger the detector determine the effective area $A_{\rm eff}\left(E_0; \,h_\mathrm{det},\,\theta,\,\phi\right)$ of the detector at a given height $h_\mathrm{det}$, and incoming primary direction $\left(\theta, \phi\right)$. 

As shown in figure \ref{fig:ANITAIVEvents}, the positions of the shower axes that would trigger the detector are located in a ring-like structure. This is a consequence of the electric field being stronger at off-axis positions corresponding to the angle $\psi_{\rm max}$ where coherence is maximal, as shown in figures \ref{fig:LDF2D} and \ref{fig:LDF2D_BInc}. Several dependencies and effects can be seen in this active area around the shower axis:

%-------------------------------------------------------------
\begin{figure}[tb]
    \centering
    \includegraphics[width = .9\textwidth]{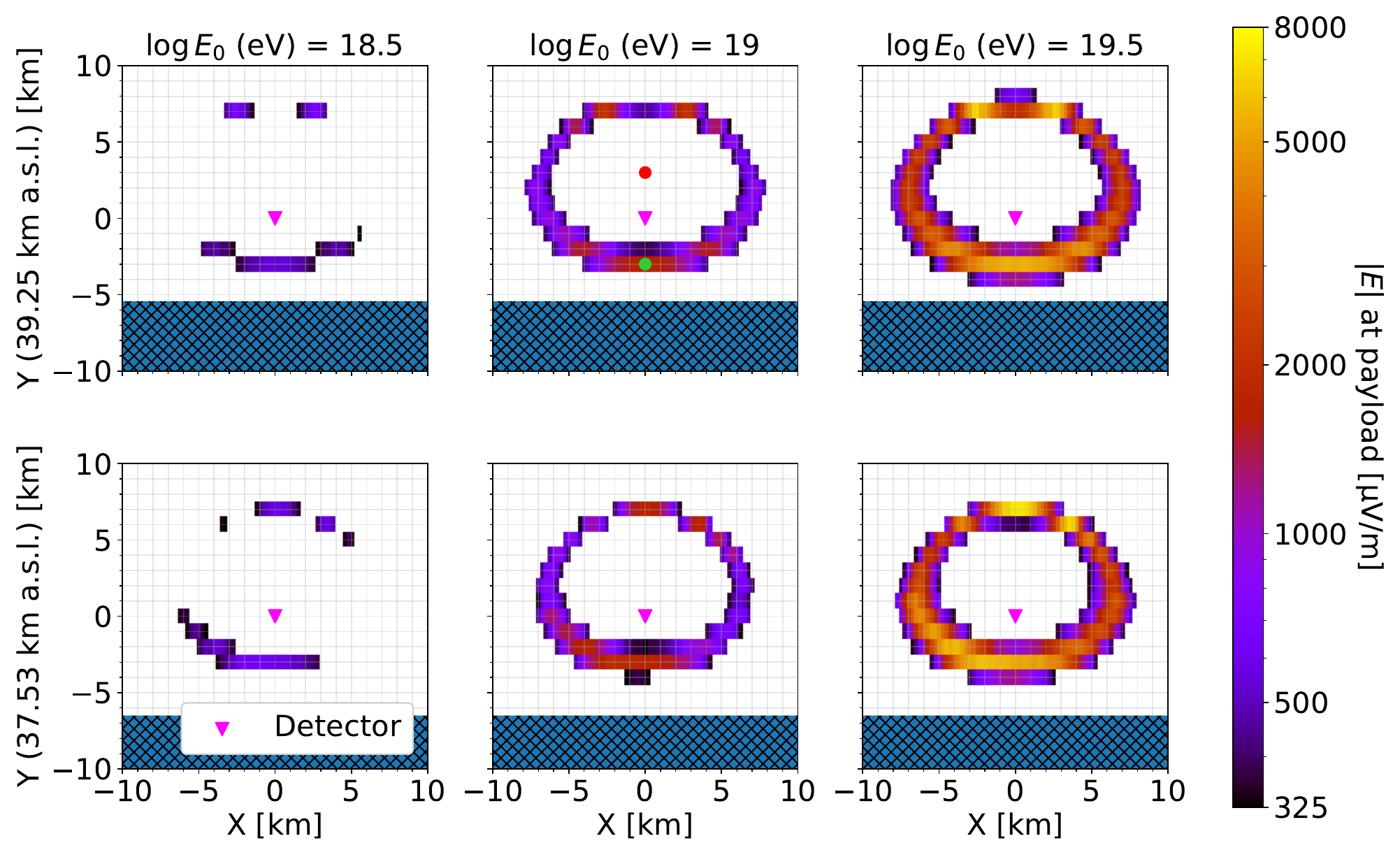}
    \caption{Active area around the position of the detector where \textit{impacting} shower axes can trigger, for the two directions $(\theta,\phi)$ of the Atmosphere-Skimming events (event “A" with ID 9734523 and $\theta_{\rm A}=95.64^\circ$ in the first row, event “B" with ID 51293223 and $\theta_{\rm B}=95.38^\circ$ in the second row) detected by ANITA IV \cite{ANITAIV}, assuming proton primaries of different energies (see text for details). The colored areas represent the positions of the shower axis in the $X-Y$ plane (perpendicular to ground and to the shower axis) of those showers that can produce a trigger at the position of the payload (magenta triangle at $(X,Y)=(0,0)$). The color scale indicates the peak electric field in the $300-1000\unit{MHz}$ frequency band induced by the shower at the  detector. The green and red dots in the middle top panel at $(X,Y)=(0,-3)\unit{km}$ and $(X,Y)=(0,3)\unit{km}$ respectively, represent the impact points on the plane of the axes of two showers, one triggering (green) and the other not triggering (red) the detector. The striped region at the bottom of each panel correspond to positions of shower axis that would hit the ground and would not be detected.} 
    \label{fig:ANITAIVEvents}
\end{figure}
%-------------------------------------------------------------

\begin{itemize}

\item
As energy increases, the region where the shower axes trigger the detector naturally becomes larger, since the threshold condition can be fulfilled increasingly further away from the viewing angle $\psi_{\rm max}$ where the amplitude of the electric field is maximal. Also, the intensity of the field, indicated by the color scale, is stronger with energy as expected.

\item
The width of the ring-like shaped structure is seen to depend on the altitude $h$ of the shower axis.
For smaller values of $h$ (lower values of $Y$ in figure \ref{fig:ANITAIVEvents}), the shower develops further away from the detector and along larger atmospheric densities. Under these conditions, the shower profiles in the longitudinal direction are shorter and narrower \cite{Tueros_Showers_JCAP}, and the time delays between different stages of the development are naturally smaller. This leads to higher signals further away from $\psi_{\rm max}$ producing a widening of the rings in figure \ref{fig:ANITAIVEvents} as $Y$ decreases.

\item
The ring-like shape of the active detector area is not centered around the position of the detector. 
Instead, its geometrical center is displaced towards higher altitudes ($Y>0\unit{km}$). This effect is due to the refractive asymmetry explained in Section \ref{sec:refractive_asym}. Since the strongest electric fields around shower axis lie in an asymmetric ring stretched towards lower altitudes, showers whose axes pass at $Y>0\unit{km}$ trigger at further distance from the detector than those passing at $Y<0\unit{km}$, as seen in figure \ref{fig:ANITAIVEvents}.
Given the similar incoming zenith angles and corresponding payload heights of the two events, the relatively large asymmetry in the shape of the active detector area is very similar in both cases. 

\item
The active area around the detector also exhibits the effects of the coherence asymmetry explained in Section \ref{sec:coherence_asym}. This reduces the high-frequency content, and thus the amplitude of the electric field at the detector, whenever the shower is observed from inside the plane containing the $\vect{v}\cross\vect{B}$ vector, where the shower front is significantly spread.
For example, in the case of event A (upper row in figure \ref{fig:ANITAIVEvents}), the incoming azimuth angle is $\phi_{\rm A}=2.01^\circ$ relative to the magnetic north, and the Lorentz force spreads the shower front in a plane almost parallel to the ground. As a consequence, air showers with axes passing above or below the position of the detector are observed from outside the plane where the shower front spreads, giving rise to stronger electric fields than those produced by showers passing through locations \textit{at the sides} of the payload.
In the case of event B (lower row in figure \ref{fig:ANITAIVEvents}), the incoming azimuth angle is $\phi_{\rm B} = 306.45^\circ$ relative to the magnetic north. Since the magnetic field at the South Pole is not perfectly vertical, the Lorentz force spreads the shower front in this case in a plane slightly tilted relative to the ground. Because of this, the axes of the showers that induce the strongest electric field will not be exactly above or below the detector, but also slightly tilted to the sides, as can be seen in the lower panels of figure \ref{fig:ANITAIVEvents}. 
The coherence asymmetry can reduce the detectable geometries and hence the active area around the detector, especially at the lowest energies. For example, in the top left panel of figure \ref{fig:ANITAIVEvents}, only AS showers that pass (approximately) above or below the detector can produce a trigger for a primary energy of $10^{18.5}\unit{eV}$. On the other hand, if the detector sees the shower \textit{from the side} (and thus falls close or even inside the plane of spreading of the shower front), the amplitude of the electric field would not be large enough to be detected.

\end{itemize}
 
The active area around the detector position also exhibits a dependence on the frequency band of operation. This is shown in figure \ref{fig:PUEO_Aeff} for the illustrative case of the arrival direction of event A, where the active area was obtained in the $50-300\unit{MHz}$ (top panels) and $300-1000\unit{MHz}$ (bottom panels) bands, corresponding to the approximate ranges of operation of the LF and MI instruments of the PUEO detector. A reduction of a factor $4$ in the trigger threshold of PUEO with respect to the $325\unit{\mu V/m}$ reference for ANITA IV was assumed in both frequency bands\footnote{This is expected to be conservative, compared to the reduction of the trigger threshold in a factor $\sim 7$ that has been estimated for the Main Instrument \cite{PUEO}.}. Some of the more salient features stemming in this figure are:

%-------------------------------------------------------------------
\begin{figure}[tb]
    \centering
    \includegraphics[width = .9\textwidth]{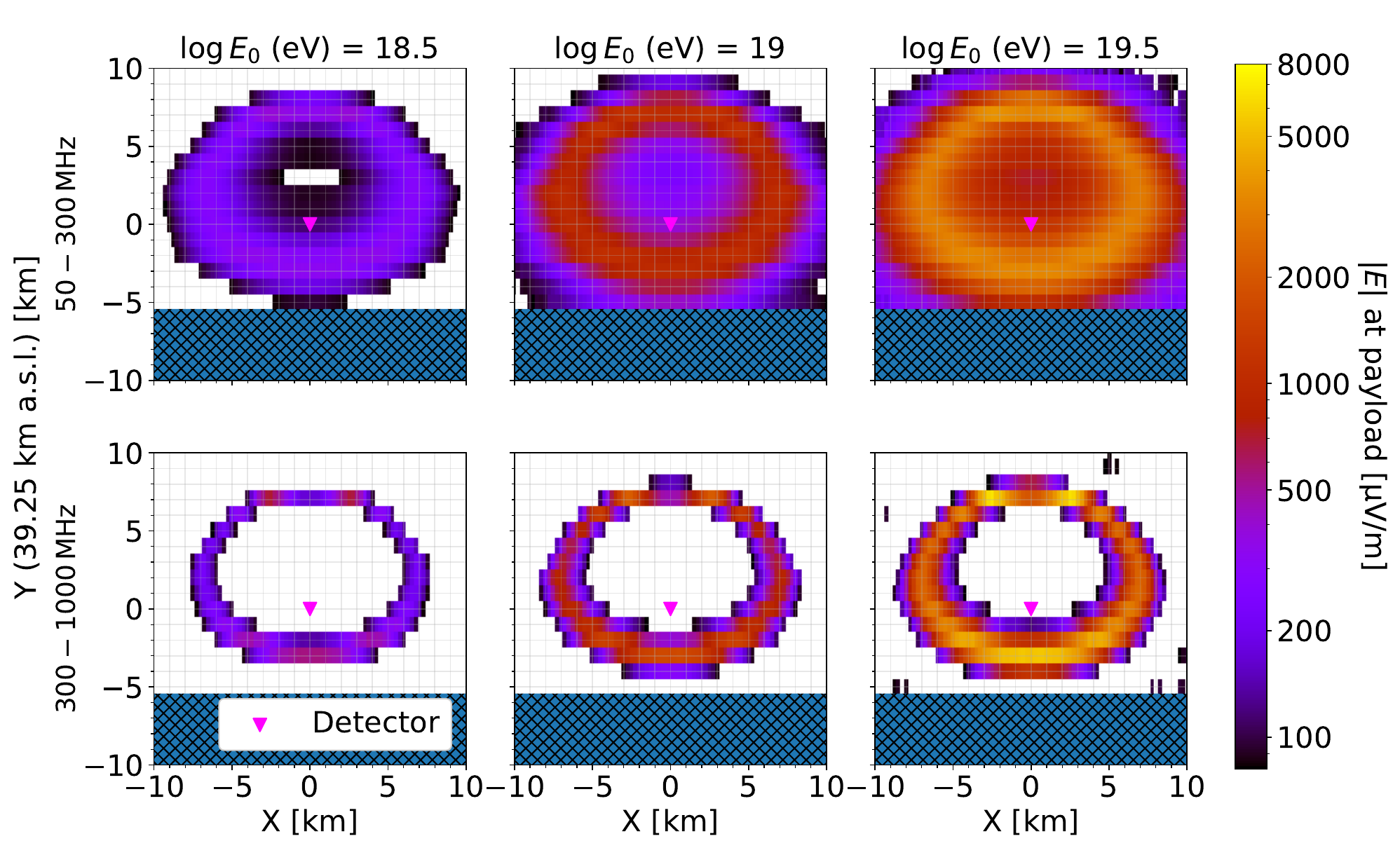} 
    \caption{Active area around the detector for cosmic ray detection at the reported incoming direction of the above-horizon event A (ID 9734523) detected by ANITA IV \cite{ANITAIV}, for the frequency bands $50-300\unit{MHz}$ (upper row) and $300-1000\unit{MHz}$ (lower row), assuming a trigger threshold four times smaller than that of ANITA IV in both frequency ranges.}
    \label{fig:PUEO_Aeff}
\end{figure}
%-------------------------------------------------------------------

\begin{itemize}

\item
As discussed in Section \ref{sec:LDF}, the lateral distribution of the electric field around the shower axis is broader in the LF band when compared to the MI band (for instance see figure \ref{fig:LDF2D}). The fall-off of the amplitude as the observer moves away from the angle $\psi_{\rm max}$ where the amplitude of the signal is maximal is slower in the LF band than in the MI range, as seen in figures \ref{fig:LDFBVer} and \ref{fig:LDFBHor}. 
The difference in the width of the peaks of the LDF translates itself into a difference in the width of the ring patterns in figure \ref{fig:PUEO_Aeff}, where it can clearly be seen that in the LF band the active area fills a larger region of the $X-Y$ plane than in the case of the MI band, to the extent that even showers whose axes pass through the detector produce a trigger in the LF band.

\item
The displacement of the ring-like pattern towards higher altitudes is similar in both frequency bands, which is consistent with the refractive asymmetry not depending on the frequency of observation as shown in Section \ref{sec:refractive_asym}. 

\item
As explained above, for the case of the particular geometry of event A, the coherence asymmetry induces a larger active area in the region above and below the detector with respect to that in the region left and right to the detector. Since this asymmetry produces a reduction of the high-frequency content of the signals (Section \ref{sec:coherence_asym}), the expectation is that it should be more prominent in the MI than in the LF band. This is seen in figure \ref{fig:PUEO_Aeff} when comparing the top and bottom panels at each energy, being somewhat more evident at low energy and in particular at $\log_{10}(E_0/\unit{eV})=18.5$.   

\item 
Regarding the trigger threshold, a reduction of it by a factor of 4 in PUEO compared to that in ANITA IV, has the effect of increasing the active area around the detector at a given energy and direction, subsequently increasing the number of expected events. This can be qualitatively checked comparing the size of the active area assuming the typical threshold for ANITA IV in the $300-1000\unit{MHz}$ band (figure \ref{fig:ANITAIVEvents}, event A, upper row), and the corresponding active area with the reduced trigger threshold peak field in PUEO (figure \ref{fig:PUEO_Aeff}, event A, lower row). The effect of the lower threshold in PUEO is more apparent at the lowest energy simulated in this work, $E_0=10^{18.5}\unit{eV}$ (compare the top leftmost panel of figure \ref{fig:ANITAIVEvents} and the bottom leftmost panel of figure \ref{fig:PUEO_Aeff}).  

\end{itemize}

The effective area of the detector $A_{\rm eff}\left(E_0; \,h_\mathrm{det},\,\theta,\,\phi\right)$ can be estimated calculating the ratio of simulated showers with primary energy $E_0$ and arrival direction $\left(\theta, \phi\right)$ that trigger a detector located at height $h_\mathrm{det}$ ($N_{\rm trig}$), relative to the number of simulated showers ($N_{\rm tot}$):
\begin{equation}
    A_{\rm eff}\left(E_0;\,h_\mathrm{det},\,\theta,\,\phi\right)\sim A_{\rm sample}\,\frac{N_{\rm trig}}{N_{\rm tot}}\,\cos\left(\theta-\frac{\pi}{2}\right)\,.
    \label{eq:Aeff}
\end{equation}
Here $A_{\rm sample}$ is the sampled area in the detector plane, and the cosine factor accounts for the projection of the effective area along the arrival direction of the showers.

\begin{figure}[tb]
    \centering
    \includegraphics[width = .85\textwidth]{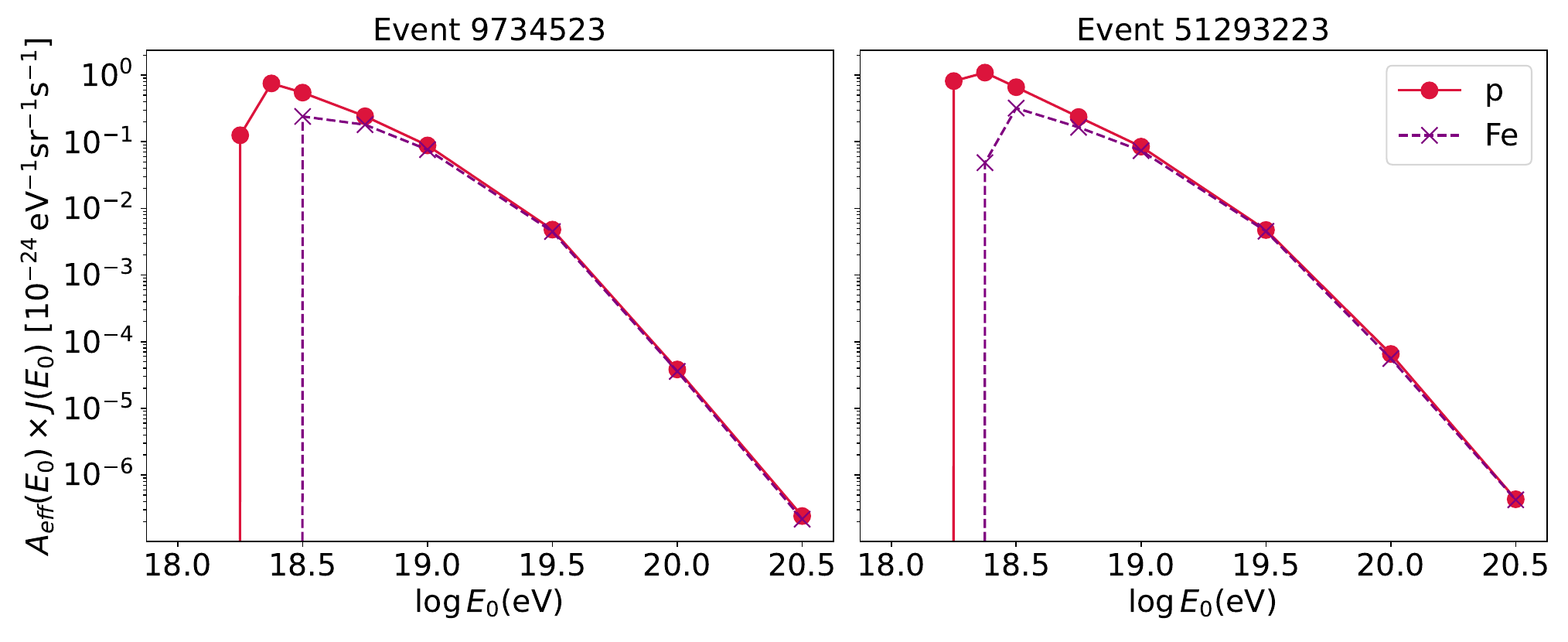}
     \caption{Estimated effective areas for the incoming directions of the two Atmosphere-Skimming events detected by ANITA IV \cite{ANITAIV} multiplied by the cosmic-ray flux as measured at the Pierre Auger Observatory \cite{PierreAuger_spectrum}. Left panel for event A and right panel for event B. Both proton and iron primaries are assumed.}
    \label{fig:ANITAIVAeff}
\end{figure}

The effective area of ANITA IV can be folded with the cosmic-ray energy spectrum to obtain the response of the detector (proportional to the number of expected events) for the particular geometries of events A and B. This is shown in figure \ref{fig:ANITAIVAeff}, either assuming proton (solid red lines) or iron (dashed blue lines) primaries. For the cosmic-ray energy spectrum we used the measurements at the Pierre Auger Observatory \cite{PierreAuger_spectrum}. The estimated effective areas would be roughly consistent with the observed number of events, given all the approximations involved. For example, the integration of the detector response in the whole range of simulated energies and incoming directions compatible with the event ``A'' (i.e., within the ANITA IV pointing resolution of $\sim 0.2^\circ$) yields an expected number of events of $N\sim 0.3$ over a live time period of $24.5$ days \cite{ANITAIV}.

%-------------------------------------------------------------
%---------------------------------------------------------------------------

The increase of the effective area with primary energy (apparent in figure \ref{fig:ANITAIVEvents}) does not compensate the rapid fall-off of the cosmic-ray flux above $\sim 10^{18.5}\unit{eV}$ inducing a peak in the response of the detector at around that energy. Given that the minimum cosmic-ray energy needed to produce a trigger is $\sim10^{18.25}\unit{eV}$, the results in figure \ref{fig:ANITAIVAeff} indicate that the most likely energy of the two AS events detected in ANITA IV could lie between $\sim1.8$ and $\sim3.2\unit{EeV}$ where the response function, proportional to the number of expected events, is maximum. 

The cosmic-ray energy above which showers trigger, and the number of expected events are both smaller for iron than for proton primaries. This arises due to a combination of two effects. For the same geometry and primary energy, the distance needed for the shower to reach its maximum decreases as the mass of the primary particle increases and the shower develops further from the detector \cite{Tueros_Showers_JCAP}. This would reduce the amplitude of the electric field at the detector in the case of iron showers. In addition, the amount of energy channeled into the electromagnetic component of the air shower, responsible for the bulk of radio emission, is smaller in the case of heavy primary particles \cite{PierreAuger_invisible, Tueros_Showers_JCAP}. 
This would also reduce the amplitude of the electric field relative to the case of proton primaries. 

%%%%%%%%%%%%%%%%%%%%%%%%%%%%%%%%%%%%%%%%%%%%%
\section{Summary and Conclusions}
\label{sec:Conclusions}

In this work, we have performed a characterisation of the properties of the radio emission in atmosphere-skimming (AS) showers using, for the first time, detailed 4D microscopic simulations of the particle cascades. The peculiar geometry of AS events (figure \ref{fig:RASPASSAntennas}) produces significant differences in the properties of the particle cascades, when compared to regular downward-going air showers, that give rise to unique features in the radio emission. 

The propagation of the radio signal across an index of refraction gradient induces an expected \textit{refractive asymmetry} (figures \ref{fig:LDFBVer} to \ref{fig:LDF2D_BInc}). The lateral distribution of the emitted radiation appears displaced towards lower altitudes as a consequence of the different optical paths traversed by the emitted radiation around the shower axis. The effect, which does not depend on frequency, is stronger for showers with smaller impact parameter $b$ (with either larger values of $\theta$ or smaller $h$), for which the optical paths above or below the shower axis are significantly different. This effect, also observed in simulations of inclined downward-going air showers, is enhanced due to the almost horizontal geometry of the shower axis in AS events.

The long distances over which AS events unfold give ample time for the Earth's magnetic field to deflect charged particles, producing a significant spread of the shower front along the direction of the Lorentz force. This asymmetry in the lateral development of the particle cascade gives rise to a unique \textit{coherence asymmetry} in the radio signal. Observers placed \textit{inside} the plane containing the $\vect{v}\times\vect{B}$ vector, where the shower front spreads, receive signals with a significantly reduced high frequency content, due to the increased time delays between the emission of different parts of the shower front. This effect therefore alters the lateral distribution of the signal around shower axis, reducing the amplitude of the signals when observed along the direction of flattening of the shower (figures \ref{fig:LDFBVer} to \ref{fig:LDF2D_BInc}), and has a strong impact on the frequency spectrum of the simulated electric fields, especially in the high-frequency range (figures \ref{fig:Spectra} and \ref{fig:Spectral_fits}). The magnitude of the coherence asymmetry depends both on the frequency band of observation and on the geometry of the shower, which will be more affected by the geomagnetic field as the impact parameter $b$ increases.

Detailed simulations are needed to address the impact of the unique features of the radio emission on the interpretation of data collected by experiments using the radio technique. The effects of the refractive and coherence asymmetries are expected to depend on the altitude, location and frequency response of the detector. This situation calls for dedicated studies, now feasible using ZHAireS-RASPASS, taking into account the design properties and science objectives of the specific detector, as well as its detection environment. Current analysis methods should also be revisited before being applied to AS events. As an example shown in figure \ref{fig:Spectral_fits}, the coherence asymmetry could give rise to relative uncertainties of the order of $\sim 20\%$ in the estimated energy of events recorded aboard balloon-borne radio detectors.

The exposure of high-altitude experiments to AS events is also affected by the refractive and coherence asymmetries. This is illustrated in figure \ref{fig:ANITAIVEvents} for the incoming directions of the two atmosphere-skimming events recorded by ANITA IV. While the refractive asymmetry displaces the ring-like shape of the effective area towards higher altitudes, the coherence asymmetry can, at energies close to the detection threshold, restrict the shower geometries that can produce a trigger to those that leave the detector outside of the plane of flattening of the shower front. The change of effective area depending on the frequency range and trigger threshold of the detector was studied in figure \ref{fig:PUEO_Aeff}, considering the frequency bands of the LF and MI instruments of PUEO. The increase in effective area in the low frequency range relative to the MI band, due to the widening of the peaks of the LDFs around their peaks (figures \ref{fig:LDFBVer} to \ref{fig:LDF2D_BInc}), is accompanied by a reduction in the magnitude of the coherence asymmetry, that has its main effect in the high-frequency regime.

\looseness=-1
The improved detection thresholds and the extension of the frequency range of operation in the next generation of balloon-borne radio experiments is expected to increase the number of registered atmosphere-skimming events with respect to the previous flights of ANITA. This situation, together with the development of current and future experiments focused on the detection of cosmic rays, photons and neutrinos of astrophysical origin, will require dedicated and detailed Monte Carlo simulations.
The flexibility of RASPASS, together with the reliability inherited from ZHAireS, can become a powerful tool for a wide range of detector types, both at the design stage and in the analysis of real data.

\section{Acknowledgments} 
This work has received financial support from
Mar\'\i a de Maeztu grant CEX2023-001318-M funded by MICIU/AEI /10.13039/501100011033;
Xunta de Galicia, Spain (CIGUS Network of Research Centers and 
Consolidaci\'on 2021 GRC GI-2033 ED431C-2021/22 and 2022 ED431F-2022/15);
Feder Funds;
Ministerio de Ciencia, Innovaci\'on y Universidades/Agencia Estatal de Investigaci\'on, Spain
(PID2019-105544GB-I00, PID2022-140510NB-I00, PCI2023-145952-2);
and European Union ERDF.

% Using bibtex package
\bibliography{main_jcap}

\providecommand{\noopsort}[1]{}\providecommand{\singleletter}[1]{#1}%

\providecommand{\href}[2]{#2}\begingroup\raggedright\begin{thebibliography}{10}

\bibitem{ANITA}
{\scshape ANITA} collaboration, \emph{{The Antarctic Impulsive Transient
  Antenna Ultra-high Energy Neutrino Detector Design, Performance, and
  Sensitivity for 2006-2007 Balloon Flight}},
  \href{https://doi.org/10.1016/j.astropartphys.2009.05.003}{\emph{Astropart.
  Phys.} {\bfseries 32} (2009) 10}
  [\href{https://arxiv.org/abs/0812.1920}{{\ttfamily 0812.1920}}].

\bibitem{ANITAI-II}
{\scshape ANITA} collaboration, \emph{{Characteristics of Four Upward-pointing
  Cosmic-ray-like Events Observed with ANITA}},
  \href{https://doi.org/10.1103/PhysRevLett.117.071101}{\emph{Phys. Rev. Lett.}
  {\bfseries 117} (2016) 071101}
  [\href{https://arxiv.org/abs/1603.05218}{{\ttfamily 1603.05218}}].

\bibitem{ANITAIII}
{\scshape ANITA} collaboration, \emph{{Observation of an Unusual Upward-going
  Cosmic-ray-like Event in the Third Flight of ANITA}},
  \href{https://doi.org/10.1103/PhysRevLett.121.161102}{\emph{Phys. Rev. Lett.}
  {\bfseries 121} (2018) 161102}
  [\href{https://arxiv.org/abs/1803.05088}{{\ttfamily 1803.05088}}].

\bibitem{ANITAIV}
{\scshape ANITA} collaboration, \emph{{Unusual Near-Horizon Cosmic-Ray-like
  Events Observed by ANITA-IV}},
  \href{https://doi.org/10.1103/PhysRevLett.126.071103}{\emph{Phys. Rev. Lett.}
  {\bfseries 126} (2021) 071103}
  [\href{https://arxiv.org/abs/2008.05690}{{\ttfamily 2008.05690}}].

\bibitem{PUEO}
{\scshape PUEO} collaboration, \emph{{The Payload for Ultrahigh Energy
  Observations (PUEO): a white paper}},
  \href{https://doi.org/10.1088/1748-0221/16/08/P08035}{\emph{JINST} {\bfseries
  16} (2021) P08035} [\href{https://arxiv.org/abs/2010.02892}{{\ttfamily
  2010.02892}}].

\bibitem{Olinto:2023vmx}
{\scshape POEMMA, JEM-EUSO} collaboration, \emph{{POEMMA (Probe Of Extreme
  Multi-Messenger Astrophysics) Roadmap Update}},
  \href{https://doi.org/10.22323/1.444.1159}{\emph{PoS} {\bfseries ICRC2023}
  (2023) 1159} [\href{https://arxiv.org/abs/2309.14561}{{\ttfamily
  2309.14561}}].

\bibitem{BEACON}
D.~Southall et~al., \emph{{Design and initial performance of the prototype for
  the BEACON instrument for detection of ultrahigh energy particles}},
  \href{https://doi.org/10.1016/j.nima.2022.167889}{\emph{Nucl. Instrum. Meth.
  A} {\bfseries 1048} (2023) 167889}
  [\href{https://arxiv.org/abs/2206.09660}{{\ttfamily 2206.09660}}].

\bibitem{GRAND}
{\scshape GRAND} collaboration, \emph{{The Giant Radio Array for Neutrino
  Detection (GRAND): Science and Design}},
  \href{https://doi.org/10.1007/s11433-018-9385-7}{\emph{Sci. China Phys. Mech.
  Astron.} {\bfseries 63} (2020) 219501}
  [\href{https://arxiv.org/abs/1810.09994}{{\ttfamily 1810.09994}}].

\bibitem{PierreAuger_Radio}
{\scshape Pierre Auger} collaboration, \emph{{Status and expected performance
  of the AugerPrime Radio Detector}},
  \href{https://doi.org/10.22323/1.444.0344}{\emph{PoS} {\bfseries ICRC2023}
  (2023) 344}.

\bibitem{PierreAuger_FD}
{\scshape Pierre Auger} collaboration, \emph{{The Fluorescence Detector of the
  Pierre Auger Observatory}},
  \href{https://doi.org/10.1016/j.nima.2010.04.023}{\emph{Nucl. Instrum. Meth.
  A} {\bfseries 620} (2010) 227}
  [\href{https://arxiv.org/abs/0907.4282}{{\ttfamily 0907.4282}}].

\bibitem{PierreAuger_upward}
{\scshape Pierre Auger} collaboration, \emph{{Constraints on upward-going air
  showers using the Pierre Auger Observatory data}},
  \href{https://doi.org/10.22323/1.444.1099}{\emph{PoS} {\bfseries ICRC2023}
  (2023) 1099}.

\bibitem{PierreAuger_HEAT}
{\scshape Pierre Auger} collaboration, \emph{{HEAT - a low energy enhancement
  of the Pierre Auger Observatory}},
  \href{https://doi.org/10.5194/astra-7-183-2011}{\emph{Astrophys. Space Sci.
  Trans.} {\bfseries 7} (2011) 183}
  [\href{https://arxiv.org/abs/1106.1329}{{\ttfamily 1106.1329}}].

\bibitem{AugerPrime}
{\scshape Pierre Auger} collaboration, \emph{{The Pierre Auger Observatory
  Upgrade - Preliminary Design Report}},
  \href{https://arxiv.org/abs/1604.03637}{{\ttfamily 1604.03637}}.

\bibitem{TelescopeArray}
{\scshape Telescope Array} collaboration, \emph{{Telescope array experiment}},
  \href{https://doi.org/10.1016/j.nuclphysbps.2007.11.002}{\emph{Nucl. Phys. B
  Proc. Suppl.} {\bfseries 175-176} (2008) 221}.

\bibitem{POEMMA}
{\scshape POEMMA} collaboration, \emph{{The POEMMA (Probe of Extreme
  Multi-Messenger Astrophysics) observatory}},
  \href{https://doi.org/10.1088/1475-7516/2021/06/007}{\emph{JCAP} {\bfseries
  06} (2021) 007} [\href{https://arxiv.org/abs/2012.07945}{{\ttfamily
  2012.07945}}].

\bibitem{Eser:2023lck}
{\scshape JEM-EUSO} collaboration, \emph{{Overview and First Results of
  EUSO-SPB2}}, \href{https://doi.org/10.22323/1.444.0397}{\emph{PoS} {\bfseries
  ICRC2023} (2023) 397} [\href{https://arxiv.org/abs/2308.15693}{{\ttfamily
  2308.15693}}].

\bibitem{Adams:2024gsj}
J.~H. Adams et~al., \emph{{The EUSO-SPB2 Fluorescence Telescope for the
  Detection of Ultra-High Energy Cosmic Rays}},
  \href{https://arxiv.org/abs/2406.13673}{{\ttfamily 2406.13673}}.

\bibitem{Cummings:2023ypo}
{\scshape JEM-EUSO} collaboration, \emph{{Analysis of above-the-limb cosmic
  rays for EUSO-SPB2}}, \href{https://doi.org/10.22323/1.444.0527}{\emph{PoS}
  {\bfseries ICRC2023} (2023) 527}
  [\href{https://arxiv.org/abs/2310.07063}{{\ttfamily 2310.07063}}].

\bibitem{Glaser:2016qso}
C.~Glaser, M.~Erdmann, J.~R. H\"orandel, T.~Huege and J.~Schulz,
  \emph{{Simulation of Radiation Energy Release in Air Showers}},
  \href{https://doi.org/10.1088/1475-7516/2016/09/024}{\emph{JCAP} {\bfseries
  09} (2016) 024} [\href{https://arxiv.org/abs/1606.01641}{{\ttfamily
  1606.01641}}].

\bibitem{Ammerman-Yebra:2023rhr}
J.~Ammerman-Yebra, J.~Alvarez-Mu\~niz and E.~Zas, \emph{{Density and magnetic
  field strength dependence of radio pulses induced by energetic air showers}},
  \href{https://doi.org/10.1088/1475-7516/2023/08/015}{\emph{JCAP} {\bfseries
  08} (2023) 015} [\href{https://arxiv.org/abs/2305.11668}{{\ttfamily
  2305.11668}}].

\bibitem{Tueros_Radio_ICRC2023}
M.~Tueros, S.~Cabana-Freire and J.~Alvarez-Mu\~niz, \emph{{Radio-Emission from
  Atmosphere-Skimming Cosmic-Ray Showers in High-Altitude Balloon-Borne
  experiments}}, \href{https://doi.org/10.22323/1.444.0349}{\emph{PoS}
  {\bfseries ICRC2023} (2023) 349}.

\bibitem{Chiche:2024yos}
S.~Chiche, C.~Zhang, F.~Schl\"uter, K.~Kotera, T.~Huege, K.~D. de~Vries et~al.,
  \emph{{Loss of Coherence and Change in Emission Physics for Radio Emission
  from Very Inclined Cosmic-Ray Air Showers}},
  \href{https://doi.org/10.1103/PhysRevLett.132.231001}{\emph{Phys. Rev. Lett.}
  {\bfseries 132} (2024) 231001}
  [\href{https://arxiv.org/abs/2404.14541}{{\ttfamily 2404.14541}}].

\bibitem{Cummings:2021bhg}
A.~Cummings, R.~Aloisio, J.~Eser and J.~Krizmanic, \emph{{Modeling the optical
  Cherenkov signals by cosmic ray extensive air showers directly observed from
  suborbital and orbital altitudes}},
  \href{https://doi.org/10.1103/PhysRevD.104.063029}{\emph{Phys. Rev. D}
  {\bfseries 104} (2021) 063029}
  [\href{https://arxiv.org/abs/2105.03255}{{\ttfamily 2105.03255}}].

\bibitem{Cocconi_magnetic}
G.~Cocconi, \emph{{Influence of the Earth´s Magnetic Field on the Extensive
  Air Showers}}, {\emph{Phys. Rev.} {\bfseries 93} (1954) 646}.

\bibitem{Tueros_RASPASS_ARENA2022}
M.~J. Tueros, \emph{{Simulation of radiopulses from Atmosphere-Skimming
  Extensive Air Showers with ZHAireS-RASPASS}},
  \href{https://doi.org/10.22323/1.424.0056}{\emph{PoS} {\bfseries ARENA2022}
  (2023) 056}.

\bibitem{Tueros_Showers_ICRC2023}
M.~Tueros, S.~Cabana-Freire and J.~Alvarez-Mu\~niz, \emph{{Characterization of
  Atmosphere-Skimming Cosmic-Ray Showers in High-Altitude Balloon-Borne
  Experiments}}, \href{https://doi.org/10.22323/1.444.0348}{\emph{PoS}
  {\bfseries ICRC2023} (2023) 348}.

\bibitem{Tueros_Showers_JCAP}
M.~Tueros, S.~Cabana-Freire and J.~Alvarez-Mu\~niz, \emph{{Characterization of
  atmosphere-skimming cosmic-ray showers in high-altitude experiments}},
  \href{https://doi.org/10.1088/1475-7516/2024/07/065}{\emph{JCAP} {\bfseries
  07} (2024) 065} [\href{https://arxiv.org/abs/2404.01239}{{\ttfamily
  2404.01239}}].

\bibitem{Zas:1991jv}
E.~Zas, F.~Halzen and T.~Stanev, \emph{{Electromagnetic pulses from high-energy
  showers: Implications for neutrino detection}},
  \href{https://doi.org/10.1103/PhysRevD.45.362}{\emph{Phys. Rev. D} {\bfseries
  45} (1992) 362}.

\bibitem{Alvarez-Muniz:2022uey}
J.~Alvarez-Mu\~niz and E.~Zas, \emph{{Progress in the Simulation and Modelling
  of Coherent Radio Pulses from Ultra High-Energy Cosmic Particles}},
  \href{https://doi.org/10.3390/universe8060297}{\emph{Universe} {\bfseries 8}
  (2022) 297}.

\bibitem{AIRES}
S.~J. Sciutto, \emph{{AIRES: A system for air shower simulations}},
  \href{https://arxiv.org/abs/astro-ph/9911331}{{\ttfamily astro-ph/9911331}}.

\bibitem{ZHAireS}
J.~Alvarez-Muniz, W.~R. Carvalho, A.~Romero-Wolf, M.~Tueros and E.~Zas,
  \emph{{Coherent Radiation from Extensive Air Showers in the Ultra-High
  Frequency Band}},
  \href{https://doi.org/10.1103/PhysRevD.86.123007}{\emph{Phys. Rev. D}
  {\bfseries 86} (2012) 123007}
  [\href{https://arxiv.org/abs/1208.0951}{{\ttfamily 1208.0951}}].

\bibitem{IGRF13}
P.~Alken et~al., \emph{{International Geomagnetic Reference Field: the
  thirteenth generation}},
  \href{https://doi.org/10.1186/s40623-020-01288-x}{\emph{Earth Planets Space}
  {\bfseries 73} (2021) 49}.

\bibitem{Kahn-Lerche:1966}
F.~D. Kahn and L.~I., \emph{{Radiation from cosmic ray air showers}},
  \href{https://doi.org/10.1098/rspa.1966.0007}{\emph{Proc. Roy. Soc. Lond. A}
  {\bfseries 289} (1966) 206}.

\bibitem{Askaryan:1962}
G.~A. Askar'yan, \emph{{Excess Negative Charge of an Electron-Photon Shower and
  its Coherent Radio Emission}}, {\emph{Soviet Physics JETP} {\bfseries 14}
  (1962) 441}.

\bibitem{PierreAuger:2014ldh}
{\scshape Pierre Auger} collaboration, \emph{{Probing the radio emission from
  air showers with polarization measurements}},
  \href{https://doi.org/10.1103/PhysRevD.89.052002}{\emph{Phys. Rev. D}
  {\bfseries 89} (2014) 052002}
  [\href{https://arxiv.org/abs/1402.3677}{{\ttfamily 1402.3677}}].

\bibitem{Schellart:2014oaa}
P.~Schellart et~al., \emph{{Polarized radio emission from extensive air showers
  measured with LOFAR}},
  \href{https://doi.org/10.1088/1475-7516/2014/10/014}{\emph{JCAP} {\bfseries
  10} (2014) 014} [\href{https://arxiv.org/abs/1406.1355}{{\ttfamily
  1406.1355}}].

\bibitem{Refractive_displacement}
F.~Schl\"uter, M.~Gottowik, T.~Huege and J.~Rautenberg, \emph{{Refractive
  displacement of the radio-emission footprint of inclined air showers
  simulated with CoREAS}},
  \href{https://doi.org/10.1140/epjc/s10052-020-8216-z}{\emph{Eur. Phys. J. C}
  {\bfseries 80} (2020) 643}
  [\href{https://arxiv.org/abs/2005.06775}{{\ttfamily 2005.06775}}].

\bibitem{Reflex}
J.~Alvarez-Mu\~niz, W.~R. Carvalho, D.~Garc\'\i{}a-Fern\'andez, H.~Schoorlemmer
  and E.~Zas, \emph{{Simulations of reflected radio signals from cosmic ray
  induced air showers}},
  \href{https://doi.org/10.1016/j.astropartphys.2014.12.005}{\emph{Astropart.
  Phys.} {\bfseries 66} (2015) 31}
  [\href{https://arxiv.org/abs/1502.02117}{{\ttfamily 1502.02117}}].

\bibitem{ANITAEnergyFlux}
H.~Schoorlemmer et~al., \emph{{Energy and Flux Measurements of Ultra-High
  Energy Cosmic Rays Observed During the First ANITA Flight}},
  \href{https://doi.org/10.1016/j.astropartphys.2016.01.001}{\emph{Astropart.
  Phys.} {\bfseries 77} (2016) 32}
  [\href{https://arxiv.org/abs/1506.05396}{{\ttfamily 1506.05396}}].

\bibitem{GDAS}
P.~Mitra et~al., \emph{{Reconstructing air shower parameters with LOFAR using
  event specific GDAS atmosphere}},
  \href{https://doi.org/10.1016/j.astropartphys.2020.102470}{\emph{Astropart.
  Phys.} {\bfseries 123} (2020) 102470}
  [\href{https://arxiv.org/abs/2006.02228}{{\ttfamily 2006.02228}}].

\bibitem{ANITAIVTrigger}
A.~Romero-Wolf et~al., \emph{{Comprehensive analysis of anomalous ANITA events
  disfavors a diffuse tau-neutrino flux origin}},
  \href{https://doi.org/10.1103/PhysRevD.99.063011}{\emph{Phys. Rev. D}
  {\bfseries 99} (2019) 063011}
  [\href{https://arxiv.org/abs/1811.07261}{{\ttfamily 1811.07261}}].

\bibitem{PierreAuger_spectrum}
{\scshape Pierre Auger} collaboration, \emph{{The energy spectrum of cosmic
  rays beyond the turn-down around $10^{17}$ eV as measured with the surface
  detector of the Pierre Auger Observatory}},
  \href{https://doi.org/10.1140/epjc/s10052-021-09700-w}{\emph{Eur. Phys. J. C}
  {\bfseries 81} (2021) 966}
  [\href{https://arxiv.org/abs/2109.13400}{{\ttfamily 2109.13400}}].

\bibitem{PierreAuger_invisible}
{\scshape Pierre Auger} collaboration, \emph{{Data-driven estimation of the
  invisible energy of cosmic ray showers with the Pierre Auger Observatory}},
  \href{https://doi.org/10.1103/PhysRevD.100.082003}{\emph{Phys. Rev. D}
  {\bfseries 100} (2019) 082003}
  [\href{https://arxiv.org/abs/1901.08040}{{\ttfamily 1901.08040}}].

\end{thebibliography}\endgroup

\end{document}